\documentclass[aps]{revtex4}

\usepackage{graphicx}
\usepackage{amsmath}
\usepackage{amsfonts}
\usepackage{mathrsfs}
\usepackage{color}
\usepackage{slashed}
\usepackage{stackrel}
\def\<{\langle}
\def\>{\rangle}

\def\beq{\begin{equation}}
\def\eeq{\end{equation}}
\def\barray{\begin{eqnarray}}
\def\earray{\end{eqnarray}}

\newcommand{\Tr}{\mathrm{Tr}}
\newcommand{\be}{\begin{equation}}
\newcommand{\ee}{\end{equation}}
\def\ba{\begin{eqnarray}}
\def\ea{\end{eqnarray}}

\font\numbers=cmss12
\font\upright=cmu10 scaled\magstep1
\def\stroke{\vrule height8pt width0.4pt depth-0.1pt}
\def\topfleck{\vrule height8pt width0.5pt depth-5.9pt}
\def\botfleck{\vrule height2pt width0.5pt depth0.1pt}
\def\Zmath{\vcenter{\hbox{\numbers\rlap{\rlap{Z}\kern
0.8pt\topfleck}\kern 2.2pt
                   \rlap Z\kern 6pt\botfleck\kern 1pt}}}
\def\Qmath{\vcenter{\hbox{\upright\rlap{\rlap{Q}\kern
                   3.8pt\stroke}\phantom{Q}}}}
\def\Nmath{\vcenter{\hbox{\upright\rlap{I}\kern 1.7pt N}}}
\def\Rmath{\vcenter{\hbox{\upright\rlap{I}\kern 1.5pt R}}}
\def\Pmath{\vcenter{\hbox{\upright\rlap{I}\kern 1.5pt P}}}
\def\Cmath{\vcenter{\hbox{\upright\rlap{\rlap{C}\kern
                   3.8pt\stroke}\phantom{C}}}}

\begin{document}

\date{\today}

\title{The Riemann zeros as energy levels of a Dirac fermion \\
 in a potential built from the  prime numbers in Rindler spacetime}

\author{Germ\'an Sierra}

\address{Instituto de F\'{\i}sica Te\'orica,
UAM-CSIC, Madrid, Spain}

\begin{abstract}
We construct a Hamiltonian  $H_R$ whose discrete spectrum contains,  in a certain limit,  the Riemann zeros.
$H_R$   is derived  from the action  of a massless Dirac  fermion living  
 in a domain  of Rindler spacetime, in 1+1 dimensions, that has a  boundary given 
   by the world line of a uniformly accelerated observer. 
   The action contains a sum of delta function potentials  that can be viewed as partially reflecting moving  mirrors.  
An appropriate choice of the accelerations of the mirrors, provide   {\em primitive}
 periodic orbits associated to the prime numbers $p$,  whose  periods, measured by the observer's clock, 
 are $\log \, p$.  Acting on the chiral components
 of the fermion $\chi_\mp$, $H_R$  becomes  the Berry-Keating Hamiltonian $\pm (x \hat{p} +  \hat{p} x)/2$,
 where $x$ is identified with the Rindler spatial coordinate and $\hat{p}$ with  the conjugate momentum. 
 The delta function potentials  give the matching conditions of the fermion wave functions on both
 sides of the mirrors. There is also a  phase shift $e^{i \vartheta}$   for the reflection of the fermions at the boundary where
 the observer sits.  The eigenvalue problem is  solved  by   transfer matrix methods
 in the  limit where the reflection amplitudes become infinitesimally small.  We find that for generic values of $\vartheta$
 the spectrum is a continuum, where the Riemann zeros are  missing, as in 
 the  adelic Connes model. However,  for some  values of $\vartheta$,  related to the phase
 of the zeta function,  the Riemann zeros appear as discrete eigenvalues immersed in the continuum. 
 We generalize  this result to the zeros of  Dirichlet $L$-functions, associated to  primitive characters,  that are encoded 
in  the reflection coefficients of the mirrors. Finally, we  show that the Hamiltonian associated to the Riemann zeros
 belongs to class AIII, or  chiral GUE,  of Random Matrix Theory. 
\end{abstract}
\maketitle

\section{Introduction}

A century ago,  P\'{o}lya and Hilbert suggested that the imaginary part of the non trivial 
zeros of the  Riemann zeta function, would be the oscillation frequencies of a physical 
system. The reality of these frequencies  will provide a proof of the celebrated Riemann hypothesis (RH)  \cite{R59}
which has deep consequences for the distribution of the prime numbers \cite{E74}-\cite{C03}. 
There are  evidences    that the Riemann zeros are  the eigenvalues of 
a quantum Hamiltonian: i) the Montgomery-Odlykzo law according to which 
the statistical distribution of the 
{\em zeros}  is given,  locally,  by the Gaussian Unitary
Ensemble (GUE) of Random Matrix Theory (RMT) \cite{M74}-\cite{Me};
ii) analogies between trace formulas  relating periods of  classical trajectories
and spectra in Quantum Chaos Theory and explicit formulas relating
prime numbers and Riemann zeros \cite{B86}-\cite{B03},  and iii) Selberg's trace formula
relating the lengths of the geodesics on a compact Riemann surface with negative
curvature and the eigenvalues of the Laplace-Beltrami operator \cite{S42,H76}. 
The picture proposed by Berry in 1986, is that the binomium 
primes/{\em zeros} is similar  to the binomium   classical/quantum for 
a dynamical chaotic system  \cite{B86}. Furthermore, it was  conjectured that 
the  classical Hamiltonian underlying the Riemann  zeros should be  
quasi-one dimensional,  breaking  time reversal symmetry
and with  isolated periodic orbits whose periods are the logarithm of the prime
numbers (see \cite{K99,B03} for reviews of this approach,  \cite{SH11,Wat} for general references and 
\cite{MSa}-\cite{DR} for introductions and historical background).

In 1999 Berry and  Keating (BK) \cite{BK99,BK99b},  and Connes \cite{C99}, 
suggested that a spectral realization of the  Riemann zeros could be achieved from the
quantization of the  simple classical Hamiltonian $H_{\rm cl}= xp$, 
where $x$ and $p$ are the position and  momentum of a particle moving in the real line.
The $xp$  Hamiltonian is one dimensional, breaks the time reversal symmetry, is integrable, not chaotic, 
with unbounded classical trajectories and its quantization yields a continuum spectrum \cite{S07a,TM07}.  
The connection found by Berry, Keating and Connes between $xp$ 
and  the Riemann zeros, was semiclassical,  and relied on two different regularizations schemes. 
Berry and Keating introduced a Planck cell regularization of the phase space imposing the
 constraints $|x|  \geq  \ell_x$ and $|p| \geq  \ell_p$
with $\ell_x \ell_p = 2 \pi \hbar$, obtaining   semiclassical energies that agree  asymptotically 
with the average location  of the Riemann zeros \cite{BK99}. In Connes's  work,  there 
is a cutoff $\Lambda$, and the constraints $|x|  \leq  \Lambda$ and $|p| \leq  \Lambda$, 
such that in the limit $\Lambda \rightarrow \infty$, the semiclassical spectrum 
becomes a continuum with 
 missing spectral lines associated to the average 
Riemann zeros \cite{C99}. The interpretation of the {\em zeros} as missing spectral  lines, 
would  also explain a mysterious  sign
problem  in the fluctuation term of the  number of zeros \cite{C99}. 
The possible connection between 
$xp$ and the Riemann zeros motivated several works in the past two decades,
some of them will be  discussed  in more detail below
%\cite{A99,S05,S07a,TM07,S07b, S08,L09,B08,ST08,ES10,RJ10,SL11,S12,BK11,S11,S11b,G12,MS12}.  
\cite{A99}-\cite{N14}. 

The previous  semiclassical versions of $xp$ were formulated
as consistent quantum mechanical models in references \cite{ST08,SL11,S12,BK11}. 
Connes's  version was  realized in terms of  a charge particle moving in a
box  of size $\Lambda  \times \Lambda$, 
and subject to the action of a uniform perpendicular magnetic field and 
an electrostatic potential $xy$ \cite{ST08}. 
For strong  magnetic fields the dynamics is restricted  to the lowest
Landau level, where the $xy$ potential acts effectively as the quantum  $xp$ Hamiltonian. 
 In this realization,   the smooth part of the counting formula of  the Riemann zeros
appears as a shift of the energy levels (that become a continuum in the limit $\Lambda \rightarrow \infty)$ 
and  not as an indication of  missing spectral lines. 
The Landau  model with $xy$ potential, has been used  in   an analogue model of Hawking radiation
in a  quantum Hall fluid \cite{St13}.

On the other hand, the  Berry-Keating  version of $xp$ was revisited recently  using  the classical 
Hamiltonian $H_{x/p}  = x (p + \ell_p^2/p)$ defined in the half line $x \geq \ell_x$ \cite{SL11,S12} (hereafter denoted as the 
$x/p$ model).  The role of the term $\ell_p^2 x/p$ is to bound the classical trajectories which become  periodic, 
unlike  the trajectories of  $xp$ that  are unbounded. The $x/p$ Hamiltonian  can be quantized
in terms of the operator $\hat{H}_{x/p} = \sqrt{x} ( \hat{p} + \ell_p^2 \hat{p}^{-1} ) \sqrt{x}$, where
$\hat{p} = - i \hbar d/dx$ and $(\hat{p}^{-1}  \psi)(x) = - i/\hbar \int_x^\infty dy \,  \psi(y)$, and its 
 spectrum  agrees asymptotically with the average
Riemann zeros  provided $\ell_x \ell_p = 2 \pi \hbar$. 
A similar result was obtained by Berry and Keating   using the Hamiltonian
$H_{BK} =  ( x + \ell_x^2/x)  (p + \ell_p^2/p)$ that is invariant under the exchange  $x \leftrightarrow p$ \cite{BK11}.
These two works provided an spectral  realization of the average Riemann zeros, but not of the actual {\em zeros}. 
From  the Quantum Chaos perspective, the reason of this failure lies in the fact that  these variants of $xp$ are non chaotic 
and does not contain periodic orbits related to the prime numbers \cite{BK11}. More generally, any one dimensional
classical and  conservative Hamiltonian is  integrable and therefore non chaotic, which seems to lead to
nowhere. 

In this work we propose a solution of the puzzle that leads to a spectral realization of the Riemann zeros. 
The main ideas can be explained as follows.  Let us consider a chaotic billiard
in two spatial dimensions, such as the Sinai's billiard \cite{S70}-\cite{St99}.
%\cite{S70,B81,BGS84,St99}. 
A ball thrown with some energy follows chaotic
trajectories that in most cases cover the entire table, except for a discrete set  of  periodic trajectories,
whose periods, that are independent of the energy, 
 dominate the path sum  that gives rise to the Gutzwiller formula for the fluctuations
of the energy levels. Unlike this,   a one dimensional billiard, made of two walls, will be 
boring  since the ball  will go back and forth periodically between the walls. 

Let us now take semitransparent  walls, so that with a certain probability
the ball passes through or bounces off. In such a  billiard the particle may follow  several
trajectories depending on the outcome at each wall. One may say  this is a quasi-one
dimensional billiard. In Sinai's billiard, or in the motion on compact Riemann surfaces,
the particle follows geodesics, which implies  that the periods of the closed orbits are
independent of the   energy. In the quasi-1D billiard, one can achieved the same  property  by choosing 
massless particles,  say photons or massless fermions, whose trajectories lie on the  light cone 
in Minkowsky space-time. The soft walls should then be viewed as semitransparent
mirrors, or beam splitters. 

The last ingredient one has to incorporate into the 
quasi-1D billiard, or rather the array of mirrors, is chaos.
In table billiards, chaos is generated by a border that defocuses the trajectories, 
and in compact  billiards, chaos  is produced  by the negative curvature of the space that separates nearby trajectories exponentially fast. 
If the 1D mirrors stay at fix positions,  nearby light-ray trajectories will stay close in space-time.
However,  if the mirrors are accelerated,  then slightly delayed  light rays, will generally have their
reflected rays departing exponentially fast from one another after several reflections. Hence in this   model, the source of chaos
 is acceleration. The simplest situation is when the mirrors are uniformly accelerated,
in which case they are  called moving mirrors in the literature of 
 Quantum Field Theory in curved spacetimes  \cite{wil}. We shall then consider an infinite array of  moving
mirrors whose accelerations,  and  reflection properties, will be used  to encode number theoretical
information.  In particular, we shall choose  the accelerations inversely  proportional to a power of  integers, 
that leads to the appearance of {\em primitive} periodic orbits whose periods are  the logarithms of the prime numbers. 
These periods are measured  by a moving observer whose acceleration 
sets the units of this magnitude. 

This model  realizes in a relativistic framework,  Berry's suggestion of associating primitive 
periodic orbits to  prime numbers \cite{B86}.  Quantum mechanically, the waves propagating in the array generate
an interference pattern that encodes the accelerations and reflection coefficients of the mirrors. Here we find several situations: 
destructive interference where the Riemann zeros are missing spectral lines as in the adelic Connes's model
and constructive interference where the Riemann zeros are point like spectrum embedded in a continuum. 
In both cases, the connection between the spectrum of the model and the Riemann zeros
involves a  limit where the reflection  amplitudes vanish asymptotically.
This  limit is  analogue to the semiclassical limit $\hbar \rightarrow 0$, that  leads  to the Gutzwiller formula, that
was the starting point of the analogies between Quantum Chaos and Number Theory. 

The paper is organized as follows. In section II we  review the basic definitions of Rindler spacetime
that describes the  geometry of the model. We formulate  the massive Dirac fermion in 
a domain of the  Rindler space-time, we find the Hamiltonian, study its relation with $xp$ 
and recover the spectrum of the $x/p$ model,  obtaining the interpretation of the 
parameters $\ell_x$ and $\ell_p$ as inverse acceleration and fermion mass. In section III
we construct an ideal array of moving mirrors with accelerations $c^2/\ell_n$ and study the
reflections of the light rays emitted and absorbed by an observer with acceleration $c^2/\ell_1$. 
Using special relativity,  we show that the proper time of the observer's clock is proportional
to $\log \ell_n$, and that the choice $\ell_n \propto \sqrt{n}$ singularizes the trajectories associated to 
the prime numbers. We also propose an array where  $\ell_n \propto e^{n/2}$, which has  regular
trajectories, and  whose discrete spectrum is proportional to the integers 
(this model will be denoted {\em harmonic}). 
In section IV we construct the Hamiltonian of a massless Dirac fermion with delta
function potentials  associated to the moving mirrors of section III.  We derive
the matching conditions for the wave functions  and show
that the corresponding Hamiltonian is self-adjoint for generic values
of the accelerations  and the  reflection coefficients. The eigenvalue
problem for the Hamiltonian is formulated using transfer matrix methods,
and in a semiclassical limit we find the conditions for the existence of  discrete eigenvalues.
In so far the construction is general, but then we analyze several examples
and make contact with  the Riemann zeta function and the Dirichlet $L-$functions. 
In section V, the identify the symmetry class of the Hamiltonian 
under time reversal (T), charge conjugation (C) and parity (P). 
In the conclusions we summarize our results and discuss future developments.
In  Appendix A we derive the 
spectrum of the {\em harmonic} model.

\section{The massive Dirac fermion in Rindler spacetime}

\subsection{Rindler spacetime}

We  start with some basic definitions and set up our conventions. The 1+1  dimensional   
Minkowski space-time  is defined by a pair of  coordinates $(x^0, x^1) \in \Rmath^2$,
a flat metric  $\eta_{\mu \nu}$,  with signature $(-1, 1)$,  and   line element  
\beq
ds^2  =   \eta_{\mu \nu} dx^\mu dx^\nu, \qquad - \eta_{00} =  \eta_{11} = 1, \quad \eta_{01} = \eta_{10} =0, 
\label{R8}
\eeq
that  is  invariant under translations and Lorentz transformations (we set  the  units of  the speed of light $c=1$). 
The change of coordinates, 
\beq
x^0 =  \rho  \,  \sinh \phi, \quad x^1 = \rho \,  \cosh \phi, 
\label{R12}
\eeq
 brings  (\ref{R8}) to  the form 
\beq
ds^2 = d \rho^2 - \rho^2 \, d \phi^2.
\label{R13}
\eeq
$\rho$  and $\phi$ are the Rindler space and time coordinates  respectively \cite{R66,Gra}. The Rindler metric (\ref{R13}) 
 is invariant under shifts of $\phi$, generated by the Killing vector $\partial/\partial \phi$. 
 Restricting $\rho$ to positive values, defines the   region called  right Rindler wedge \cite{R66}-\cite{URMP}, 
 \beq 
{\cal R}_+  = \left\{ (\rho, \phi) \;   | \;\rho  >    0, \; - \infty <  \phi <  \infty \right\}, 
\label{R140}
\eeq
that in Minkowski coordinates is described by  the quadrant  $x^1 > |x^0|$. 
The value  $\rho=0$ represents a horizon of the  metric (\ref{R13}),
that is similar to the horizon of a  black hole, the reason being 
that the  Schwarzschild metric, near the black hole  horizon, is approximately the Rindler metric \cite{Gra}.

\begin{figure}[t]
\begin{center}
\includegraphics[height= 6.0 cm]{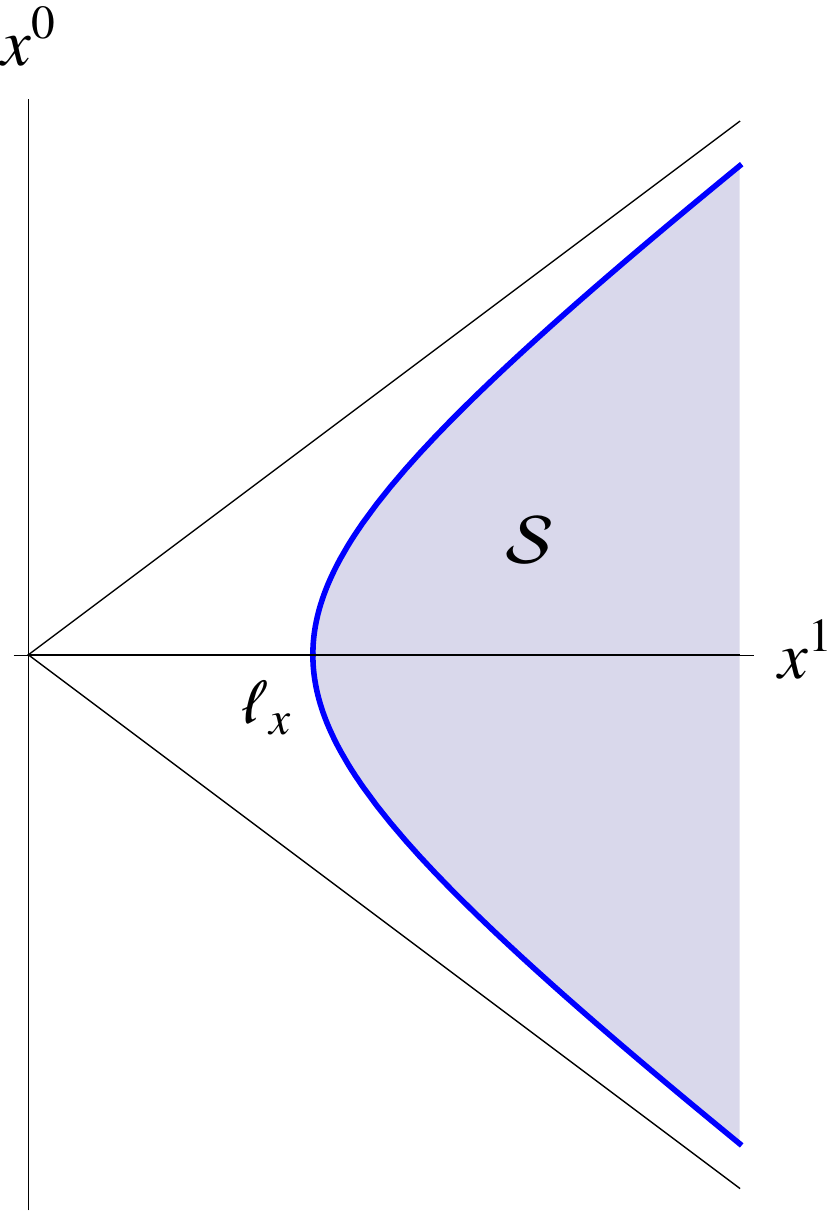} \hspace{1 cm}
\includegraphics[height= 6.0 cm]{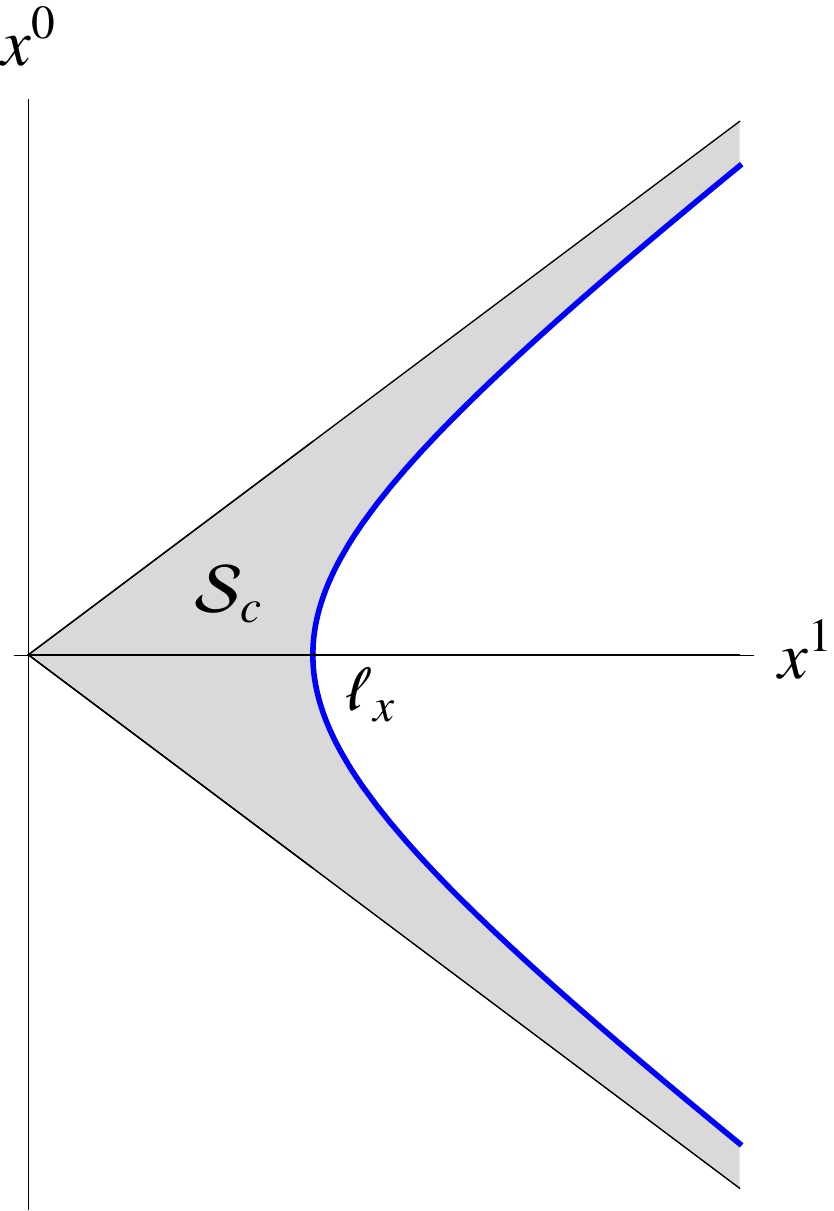}
\end{center}
\caption{ Regions  ${\cal S}$  and ${\cal S}_c$ defined in eq. (\ref{R14}). 
 The union of ${\cal S}$ and ${\cal S}_c$ is the right Rindler wedge (\ref{R140}). 
}
\label{S-space}
\end{figure}

Rindler spacetime  is  the natural arena to study  
the physical phenomena  associated to  accelerated observers \cite{R66}, such as the Unruh effect \cite{U76,URMP}. 
Let us consider an observer whose  world line is given by (\ref{R12}), 
 with $\rho = \ell >0$ a constant value. The proper time, $\tau$, 
 measured by the observer is defined by $ds^2 = - d \tau^2$, and
 its relation  to $\phi$ follows from (\ref{R13})
 \beq
 \tau = \ell \, \phi.
 \label{R15}
 \eeq
 plus  an additive constant that is set to 0 in (\ref{R15}). 
 The observer's trajectory (\ref{R12}),  written in terms of its proper time, 
 reads  
 \beq
x^0 =  \ell   \,  \sinh ( \tau/\ell) , \quad x^1 = \ell  \,  \cosh( \tau/\ell), 
\label{R16}
\eeq
and has a constant proper acceleration, $a$,  defined as the Minkowski norm 
of the vector   $a^\mu$
\beq
a^\mu = \frac{ d^2 x^\mu}{ d \tau^2}, \qquad a^\mu a_\mu = - a^2  \rightarrow  a = \frac{1}{\ell}. 
\label{R17}
\eeq
Restating the speed of light,  $a = c^2/\ell$. This quantity plays an central  role in the Unruh
effect according to which an observer,  with proper acceleration $a$, detects a thermal
bath with temperature $T_U = a/(2 \pi)$ (in units $c= \hbar = T_B=1$) \cite{U76}. Replacing $a$
by the surface gravity $\kappa$ of an observer near the horizon of a black hole, yields
the Hawking temperature $T_H = \kappa/( 2 \pi)$ \cite{H74,H75}. The similarity between the Hawking
and Unruh formulas lies in the equivalence principle of General Relativity. 

In our approach to a   spectral realization of the Riemann zeros, we shall 
introduce an observer with acceleration, $a_1=1/\ell_1$. 
The observer's world line, $\rho = \ell_1$,  divides 
${\cal R}_+$ into the   regions  ${\cal S}$   and ${\cal S}_c$
located to her  right and  left, 
 \barray 
{\cal S} &  = &  \left\{ (\rho, \phi) \;   | \;\rho  \geq \ell_1, \quad - \infty <  \phi <  \infty,  \right\},
\label{R14} \\
{\cal S}_c  & = &  \left\{ (\rho, \phi) \;   |  \; 0 <   \rho  \leq \ell_1, \quad - \infty <  \phi <  \infty,  \right\},
\nonumber 
\earray
such that ${\cal R}_+ = {\cal S}  \cup {\cal S}_c$. 
%Later on, we shall construct 
%a model associated to the Riemann zeta function, and another  one associated to the
%integers ({\em harmonic} model). In the {\em zeta}  model, $\ell_1$ will be associated
%to the number 1, while in the harmonic model,  $\ell_1$ will be associated to 0, in which case it  will be 
%denoted as   $\ell_0$. 
Let us next study the dynamics of a Dirac fermion in  ${\cal  S}$.

\subsection{Dirac fermion}

A  representation of the Dirac's gamma matrices in 1+1 dimensions  is given by \cite{yellow} 
\beq
\gamma^0 = \sigma^x,  \quad 
\gamma^1 = - i \sigma^y,  \quad \gamma^{5} \equiv  \gamma^0 \gamma^1 = \sigma^z,  
\label{R32}
\eeq
where $\sigma^a \; (a=x,y,z)$ are the Pauli matrices.  $\gamma^\mu$ satisfy 
the Clifford algebra 
\beq
\left\{ \gamma^\mu, \gamma^\nu  \right\} = - 2 \eta^{\mu \nu}, \qquad \left\{ \gamma^\mu, \gamma^5 \right\} = 0, \qquad  (\mu=0,1). 
\label{R33}
\eeq
In an abuse of notation, $\gamma^5$  denotes   the 1+1 analogue of the 1+3
gamma matrix  $\gamma^5 = \gamma^0 \gamma^1 \gamma^2 \gamma^3$, and it defines
the chirality of the fermions. 
A  Dirac fermion $\psi$  is a two component spinor 
\beq
\psi = \left( 
\begin{array}{c}
\psi_- \\
\psi_+ \\
\end{array}
\right),  \qquad \bar{\psi} = \psi^\dagger \gamma^0 = ( \psi^\dagger_+, \psi^\dagger_-), 
\label{R34}
\eeq
where ${\psi}^\dagger_\pm$ are  the conjugate of $\psi_\pm$. The fields $\psi_\mp$ are the chiral components of $\psi$,
namely  $\gamma^5 \psi_\mp = \pm \psi_\mp$.
Let us introduce the
light cone coordinates $x^\pm$,  and the  derivatives $\partial_\pm = \partial/\partial {x^\pm}$ 
\barray 
x^\pm & =  & x^0 \pm x^1 = \pm \rho \, e^{ \pm \phi},  \label{R35} \\
\partial_\pm & = & \frac{1}{2} ( \partial_0 \pm \partial_1) = 
 \pm \frac{1}{2} e^{ \mp \phi} ( \partial_\rho \pm \rho^{-1} \partial_\phi),  
\nonumber 
\earray 
where the Minkowski  metric (\ref{R8}) becomes  $ds^2 = - dx^+ dx^-$. 
Under a Lorentz  transformation with boost parameter $\lambda$, that is velocity $v = \tanh \lambda$, the 
light cone coordinates and the Dirac spinors   transform as
\beq
x^\pm \rightarrow e^{\mp \lambda} \, x^\pm, \qquad   \partial_\pm \rightarrow e^{\pm \lambda} \, \partial_\pm, \qquad
\psi_\pm \rightarrow e^{\pm  \lambda/2} \, \psi_\pm,
\label{R351}
\eeq
and  the Rindler coordinates  as 
\beq
\phi \rightarrow \phi - \lambda, \qquad \rho \rightarrow \rho.
\label{R352}
\eeq
Hence the spinors $\chi_\pm$ 
\beq
\chi_\pm = e^{ \pm \phi/2} \, \psi_\pm,
\label{R40}
\eeq
remain    invariant under (\ref{R352}).  The  spaces   ${\cal S}, {\cal S}_c$ and ${\cal R}_+$
are mapped into themselves  under Lorenzt transformations.

\subsection{Dirac action  of a massive fermion}

The Dirac action of  a fermion with mass $m$ in the space-time  domain ${\cal S}$ is (units $\hbar = c=1$)
\barray 
S  & = &  \frac{i}{2} \int_{\cal S} d^2 x   \, \bar{\psi} ( {\slashed \partial}  + i  m) \psi, 
\label{R31}  \\
& = & \frac{i}{2} \int_{\cal S} dx^+ dx^- \left[ \psi^\dagger_- \partial_+ \psi_-  +  \psi^\dagger_+ \partial_- \psi_+    +
\frac{i m}{2}  \left(   \psi^\dagger_-  \psi_+   +  \psi^\dagger_+ \psi_-   \right) 
\right] \nonumber  \\
& = & \frac{i}{2}
\int_{- \infty}^\infty d \phi \, \int_{\ell_1}^{\infty} d \,   \rho \left[ \chi^\dagger_- (  \partial_\phi + \rho \partial_\rho  + \frac{1}{2} )  \chi_-    
+  \chi^\dagger_+ ( \partial_\phi -  \rho \partial_\rho  -   \frac{1}{2} )  \chi_+  \right.
 \nonumber  \\
& & + \left. i m  \rho \left(  \chi^\dagger_-  \chi_+   +  \chi^\dagger_+ \chi_-   \right) 
\right] . \nonumber 
\earray 

\noindent 
${\cal S}$ has a boundary $\partial {\cal S}$ corresponding to the worldline $\rho= \ell_1>0$. 
%Later on, we shall study the Dirac  model in  ${\cal R}_+$. 
The variational principle applied to  (\ref{R31}) gives the Dirac equation
\beq
( {\slashed \partial}  + i  m) \psi(x) = 0, \quad x \in {\cal S},
\label{R36}
\eeq
and  the boundary condition
\beq 
\varepsilon_{\mu \nu} \, \dot{x}^\mu \,  \bar{\psi}(x)  \gamma^\nu \delta \psi(x) = 0, \quad x \in \partial {\cal S}, 
\label{R37}
\eeq
where $\delta \psi$ is an  infinitesimal variation of $\psi$, 
$\varepsilon_{\mu \nu}$ is the Levi-Civita tensor $(\varepsilon_{01} = - \varepsilon_{10} =1, \varepsilon_{\mu \mu}=0)$, 
and $\dot{x}^\mu = d x^\mu(\ell_1, \phi)/d \phi$ is the tangent to  $ \partial {\cal S}$  in  the Rindler
coordinates (\ref{R12}). The Dirac equation (\ref{R36}) reads in components 
\beq
( \partial_0 \mp    \partial_1 ) \psi_\pm  + i  m \psi_\mp  =  0 \quad \longrightarrow  \quad \partial_\mp \psi_{\pm} + \frac{i m}{2} \psi_\mp = 0.
\label{R38} 
\eeq
In the massless case,  the fields  $\psi_\pm$ decouple and describe a
right moving fermion, $\psi_+(x^+)$,  and a  left moving fermion, $\psi_-(x^-)$, 
in terms of which one can construct a Conformal Field Theory
with central charge $c=1$ \cite{yellow}. The mass term couples the two modes 
and therefore conformal invariance is lost.  The action principle applied to the last expression of eq.(\ref{R31}) gives 
\beq
 ( \partial_\phi  \pm  \rho  \partial_\rho \pm \frac{1}{2} ) \chi_\mp  +i m  \rho  \chi_\pm = 0, 
\label{R41}
\eeq
and the boundary condition 
\barray 
 \chi_-^\dagger (\ell_1, \phi)  \;  \delta \chi_-(\ell_1, \phi)   =     \chi_+^\dagger (\ell_1, \phi)  \;  
\delta \chi_+(\ell_1, \phi) , \quad \forall \phi. 
\label{R49}
\earray 
Eqs.(\ref{R41}) and (\ref{R49})   are  of course  equivalent to (\ref{R38}) and (\ref{R37})
respectively. The infinitesimal generator of translations of the Rindler time $\phi$, 
acting on the fermion wave functions,  is the Rindler Hamiltonian $H_R$, that 
can be read off from (\ref{R41})
\beq
i \partial_\phi \chi = H_R \, \chi, \qquad \chi  = \left( 
\begin{array}{c}
\chi_- \\
\chi_+ \\
\end{array}
\right), 
\label{R42}
\eeq

\beq
H_R = 
\left( \begin{array}{cc}
- i ( \rho \,   \partial_\rho + \frac{1}{2} )  & m \rho \\ 
m \rho &  i ( \rho \,   \partial_\rho + \frac{1}{2} ) \\
\end{array}
\right) = \sqrt{\rho} \, \hat{p}_\rho \, \sqrt{\rho} \, \sigma^z + m \rho \,  \sigma^x , 
\label{R43}
\eeq
where  $\hat{p}_\rho = - i \partial / \partial \rho$, is the momentum operator associated
to the radial coordinate $\rho$. The operator 

\beq
H_{xp} = - i ( \rho \,   \partial_\rho + \frac{1}{2} ) = \frac{1}{2} ( \rho \, \hat{p}_\rho + \hat{p}_\rho \rho) =
\sqrt{\rho} \, \hat{p}_\rho \, \sqrt{\rho}, 
\label{R43b}
\eeq
coincides with the quantization of the classical $xp$ Hamiltonian proposed
by Berry and Keating \cite{BK99},  where $x$ is the radial Rindler coordinate. 
The  eigenfunctions  of (\ref{R43b}) are 

\beq
H_{xp} \, \psi_E = E \, \psi_E, \qquad \psi_E = \frac{1 }{ \sqrt{2 \pi}} \rho^{- 1/2 + i E}, 
\label{Rxpb}
\eeq
with eigenvalues $E$ in the real numbers  $\Rmath$ if   $\rho >0$ \cite{S07a,TM07}. 
Thus $H_R$ consists of two copies of $xp$,  with different signs corresponding to 
opposite chiralities that are coupled by the  mass term. In  CFT
the operator $H_R$, with $m=0$,  corresponds to the sum of Virasoro operators $L_0 + \bar{L}_0$
that generate  the dilation  transformations.

\subsection{Self-adjointess of $H_R$} 

The action (\ref{R31}) is invariant under the $U(1)$ transformation $\psi \rightarrow e^{ i \alpha} \psi$. 
The corresponding Noether current is  $J^\mu = \bar{\psi} \gamma^\mu \psi$, and 
is conserved, i.e.  $\partial_\mu J^\mu =0$. The charge associated to $J^\mu$ can be integrated
along the  space-like line   (\ref{R12}) with  constant $\phi$. Using  (\ref{R40}) one finds

\beq
\int_{\phi = {\rm cte}} dx^\mu \epsilon_{\nu \mu} \,  \bar{\psi} \gamma^\nu \psi = \int d \rho 
  \left(  \chi^*_{-} \chi_{-}   +  \chi^*_{+} \chi_{+}   \right), 
\label{no}
\eeq
where $\chi^*_\pm$ are the complex conjugate of $\chi_\pm$. 
Hence the scalar product of two wave functions, in the domain ${\cal S}$,  can  be defined as 

\beq
\langle \chi_1 | \chi_2 \rangle = \int_{\ell_1}^\infty  d \rho 
  \left(  \chi^*_{1,-} \chi_{2,-}   +  \chi^*_{1,+} \chi_{2,+}   \right). 
  \label{R431}
  \eeq
The  Hamiltonian $H_R$ is hermitean (or symmetric) respect to this scalar product if 
\barray 
\langle \chi_1 | H_R    \chi_2 \rangle  =  \langle H_R  \chi_1 |\chi_2 \rangle, 
\label{R432}
\earray 
when  acting on a subspace of the total Hilbert space. 
Partial integration gives  
  \barray 
\langle \chi_1 | H_R    \chi_2 \rangle  - \langle H_R  \chi_1 |\chi_2 \rangle = 
 \int_{\ell_1}^\infty  d \rho  \partial_\rho 
  \left[  \rho \left(  \chi^*_{1,+} \chi_{2,-}   -  \chi^*_{1,-} \chi_{2,+}  \right) \right].
  \label{R4321}
  \earray  
 Hence $H_R$ is hermitean provided 

 \beq
 \lim_{\rho \rightarrow  \infty} \rho^{1/2} \chi_\pm(\rho)= 0, \qquad 
  \chi^*_{1,-} \chi_{2,-}   -  \chi^*_{1,+} \chi_{2,+} = 0, \quad {\rm at} \; \;  \rho = \ell_1.
  \label{R433}
  \eeq
The latter   condition   is  equivalent  to eq.(\ref{R49}). The solution of (\ref{R433}) is 
\barray 
- i e^{i \vartheta}  \,     \chi_-    =       \chi_+  \quad {\rm at} \; \;  \rho = \ell_1,
\label{R50}
\earray
where $\vartheta \in [0, 2 \pi)$. The quantity $- i e^{i \vartheta}$  has the physical meaning of 
the phase shift produced by the reflection of the fermion with the boundary.  It will play a very important
role in what follows. 
%Notice that this  boundary condition
%is preserved under the gauge transformations (\ref{R314}) with  $\alpha_+ = \alpha_-$, 
%so the gauge group of the model  is now the diagonal part of $U_+(1) \otimes U_-(1)$, that is $U(1)$. 

$H_R$ is not only hermitean but also  self-adjoint. According to a theorem due to von Neumann,
an operator $H$ is self-adjoint if the deficiency  indices $n_\pm$ 
are equal \cite{vN,GP90}. These indices are   the number of linearly independent eigenfunctions
of $H$ with positive and negative imaginary eigenvalues, 
\beq 
n_\pm  = \dim \, \{ \psi_\pm   | \,  H \,    \psi_\pm =   \pm  z \psi_\pm , \,  {\rm Im} \; z>0 \}.
\label{R501}
\eeq
If  $n_+ = n_-  = 0$, the operator $H$ is essentially self-adjoint, while if 
 $n_+ = n_-  > 0$, then $H$  admits  self-adjoint extensions.
%parameterized by the unitary group $U(n_+)$. 
In the case of $H_R$, one finds $n_+=n_-=1$, 
and  the self-adjoint extensions  are parameterized by the  phase $e^{i \vartheta} $ in eq.(\ref{R50}).

\subsection{Spectrum of $H_R$  } 
  
The eigenvalues and eigenvectors of the  Hamiltonian (\ref{R43}), are given by the solutions of the 
Schroedinger equation 
\beq
H_R \,  \chi = E \, \chi, \qquad 
 \chi_\pm(\rho, \phi)  = e^{ - i E \phi}   f_\pm(\rho), \qquad  \rho \geq \ell_1, 
\label{R44}
\eeq
that satisfy  the boundary condition (\ref{R50}). $E$ is  the Rindler energy.  
The equations for   $f_\pm(\rho)$  that follows from (\ref{R43}) are
\beq
\left( \rho  \partial_\rho +  \frac{1}{2}  \pm i E   \right) f_\pm Ð \mp  i m   \rho   f_\mp = 0, 
\label{R45}
\eeq
that  lead  to the second order differential eqs. 
\beq
\left[ {\rho}^2 \, \frac{\partial^2}{\partial  {\rho}^2}  +  {\rho} \, \frac{\partial}{ \partial  {\rho}} 
- \left(  \frac{1}{2} \pm  i E \right)^2   - {m}^2 {\rho}^{2} \right] f_\pm =0, 
\label{R502}
\eeq
 whose general solution is a linear combination of  the modified Bessel functions \cite{AS72}
\beq 
 f_\pm(\rho)  = C_{1} \,  e^{ \mp i \pi/4}   \,  K_{\frac{1}{2} \pm i E} (m \rho) \pm   C_{2}  \,  e^{ \mp i \pi/4}  \, I_{\frac{1}{2} \pm i E } (m \rho).
 \label{503}
 \eeq
 The phases $e^{\pm i \pi/4}$ follows from  (\ref{R45}). Let us compute the deficiency indices of $H_R$,  for
 which it is enough  to take   $E = \pm i/2$, 
 
 \barray
 E = \frac{i}{2} & \rightarrow & 
 \left\{ \begin{array}{cl}
  f_+(\rho)  =  &  C_{1} \,  e^{ - i \pi/4}   \,  K_{0} (m \rho) +   C_{2}  \,  e^{ - i \pi/4}  \, I_{0 } (m \rho), \\ 
   f_-(\rho)  = &   C_{1} \,  e^{  i \pi/4}   \,  K_{1} (m \rho) -  C_{2}  \,  e^{  i \pi/4}  \, I_{1 } (m \rho),
 \end{array}
 \right. 
 \label{504}  \\
  E = - \frac{i}{2} & \rightarrow & 
 \left\{ \begin{array}{cl}
  f_+(\rho)  =  & C_{1} \,  e^{ - i \pi/4}   \,  K_{1} (m \rho) +   C_{2}  \,  e^{ - i \pi/4}  \, I_{1 } (m \rho), \\ 
   f_-(\rho)  = &  C_{1} \,  e^{  i \pi/4}   \,  K_{0} (m \rho) -   C_{2}  \,  e^{  i \pi/4}  \, I_{0 } (m \rho).
 \end{array}
 \right. 
 \earray 
 The functions $I_{0,1}(m \rho)$ diverge exponentially as $\rho \rightarrow \infty$, that forces 
 $C_2=0$, while  $K_{0,1}(m \rho)$
 decreases  exponentially  and are normalizable provided $\ell_1 >0$. 
 The deficiency indices are therefore equal,  $n_+ = n_-=1$ and the eigenfunctions  are (setting  $C_1=1$)
\beq
f_\pm(\rho) = e^{ \mp i \pi/4} K_{\frac{ 1}{2}  \pm i E}(m \rho) ,  \qquad \rho \geq \ell_1, 
\label{R46}
\eeq
which yields 
\beq
 \chi_\pm(\rho, \phi)  = e^{ - i E \phi  \, \mp i \pi/4}   K_{\frac{ 1}{2}  \pm i E}(m \rho),
\label{R47}
\eeq
up to a common normalization constant. 
Plugging (\ref{R47}) into (\ref{R50})
yields the equation for the eigenenergies 
\beq
e^{ i \vartheta} \, K_{ \frac{1}{2} - i E} ( m \ell_1) - K_{ \frac{1}{2} + i E} ( m \ell_1) = 0.
\label{R51}
\eeq
This equation  has positive and negative solutions, but only if  $\vartheta =0$ or $\pi$, they come 
in  pairs $\{ E_n, - E_n \}$. Moreover, if   $\vartheta =0$, then  $E_0=0$ is also an  eigenvalue. 
%Later on, we shall explain  these  properties of the spectrum   in terms of  charge conjugation symmetry of the Hamiltonian $H_R$. 
The imaginary part of the Riemann zeros  also form pairs $\{ t_n, - t_n \}$ (i.e.  $\zeta(1/2 \pm i t_n)=0$),  and 
$t=0$ is not a {\em zero}  since $\zeta(1/2) \neq 0$. This situation corresponds to the choice
$\vartheta = \pi$.  

The number of eigenvalues, $n(E)$,  in the interval  $[0, E]$ (with $E>0$) is given in the asymptotic limit 
 $E \gg  m \ell_1$ by  
\beq
 n(E)  \simeq \frac{ E}{   \pi} \left(  \log \frac{2 E}{  m \ell_1 } - 1 \right)  - \frac{\vartheta}{ 2 \pi} + O(E^{-1}), \quad E \gg  m \ell_1.
\label{R52b}
\eeq
For  $E<0$ there is a similar formula with $\vartheta \rightarrow - \vartheta$. Let us compare this  expression
with the   Riemann-Mangoldt  formula that counts the  number of  zeros, ${\cal N}(t)$,  of the zeta function $\zeta(s)$ 
that lie  in the rectangle   $0 <  {\rm Re}  \, s <1, \; 0 < {\rm Im}   \, s < t $ \cite{E74} 
\barray
{\cal N}(t) & = & \langle {\cal N}(t) \rangle   + {\cal N}_{\rm fl}(t)  \label{ze1}   \\
 \langle {\cal N}(t) \rangle &  = &  \frac{ \theta(t)}{\pi} + 1 \stackrel{t \rightarrow \infty}{\longrightarrow}  \frac{t}{ 2 \pi} \left( \log \frac{ t}{ 2 \pi} -1 \right) + \frac{7}{8} 
 \nonumber 
 \\
{\cal N}_{\rm fl}(t) & = &   \frac{1}{\pi} {\rm Im} \,  \log \zeta \left( 
\frac{1}{2} + i t \right)   = O(\log t) 
\nonumber 
\earray
where $\langle {\cal N}(t) \rangle$ is the average term and  ${\cal N}_{\rm fl}(t)$  the oscillation term. 
This  expression  agrees  with (\ref{R52b}), to order $t \log t$ and $t$,  with the identifications 
\beq
E=  \frac{t}{2} , \qquad  m \ell_1   = 2 \pi. 
\label{R52c}
\eeq
However,  the constant term $7/8$ is not  reproduced by eq.(\ref{R52b}) for $\vartheta = \pi$,
which is the choice for the absence of the zero mode $E_0=0$ (note that neither $\vartheta=0$ gives the 7/8). 

\vspace{0.5 cm}

{\bf Comments:}

\begin{itemize}

\item In the limit $m \ell_1 \rightarrow 0$, the spectrum (\ref{R52b}) becomes a continuum. 
This situation arises in three  cases: 1)  if $m >0$ is kept constant and   $\ell_1 \rightarrow 0$,   the domain ${\cal S}$ 
becomes ${\cal R}_+$, and  one recovers  the spectrum of a  massive Dirac equation in ${\cal R}_+$; 
2) if $\ell_1>0$ is kept constant, and  $m \rightarrow 0$,  the particles become massless
and the effect of the boundary $\partial {\cal S}$, is to  exchange left and right moving fermions, that constitutes a
 boundary CFT \cite{yellow}, and 3) $m$  and  $\ell_1 \rightarrow 0$ that corresponds to a massless fermion  in ${\cal R}_+$.

\item Equation (\ref{R51}) coincides with  the spectrum
of the quantum $x/p$ Hamiltonian  \cite{SL11,S12}
\beq
H_{x/p} = \sqrt{x} \left(  \hat{p} +  \ell_p^2 \,  \hat{p}^{-1}  \right)  \sqrt{x}, \qquad x \geq \ell_x, 
\label{R522}
\eeq
with the identifications 
\beq
E = \frac{E_{x/p}}{2}, \qquad m = \ell_p, \qquad \ell_1 = \ell_x.
\label{R523}
\eeq 
The origin of this coincidence lies in the fact that a  classical Hamiltonian of the form
$H = U(x) p + V(x)/p$ \cite{S12} can be formulated as a massive Dirac model
in a space-time metric built from the potentials $U(x)$ and $V(x)$ \cite{MS12}. 
In the case of the $x/p$  model this metric is flat, which allows us to formulate  this
model in terms of  the  Dirac equation in the domain ${\cal S}$. 
%In the case of the $x/p$ model this metric is flat, which leads to present formulation
%of the $x/p$ model. 

\item Gupta et al. proposed recently  the Hamiltonian $( x {\slashed p} + {\slashed p} x)/2$ 
as a Dirac variant of the $xp$ Hamiltonian \cite{G12}. The former Hamiltonian is defined in 2 spatial dimensions
and after compactification of  one coordinate  becomes 1D, with an spectrum  that depends
on a regularization parameter and which is similar to the one found  from the Landau theory
with electrostatic potential $xy$
 \cite{ST08}.

\item  Burnol has studied the causal propagation of a massive boson and a massive  Dirac fermion in
the Rindler right edge  ${\cal R}_+$ relating the scattering from the past light cone to the future light cone
to the Hankel transform of zero order and suggesting a possible relation to the zeta function \cite{burnol}.

\end{itemize}

\subsection{General Dirac action} 

For later purposes we shall introduce  
the   general relativistic invariant Dirac action in 1+1 dimensions 

\barray 
S  & = &  \frac{i}{2} \int_{\cal S}   d^2 x   \, \bar{\psi} ( {\slashed \partial}   + i {\slashed A}
 +  i  m - m'  \gamma^5) \psi, 
\label{R311} 
\earray 
where, in  addition to the mass term, 
 there is a chiral interaction $\bar{\psi} \gamma^5 \psi$ and a minimal coupling 
$\bar{\psi} {\slashed A} \psi$ to a vector potential $A_\mu$ (${\slashed A} = \gamma^\mu \, A_\mu$).  
Moreover $m, m'$ can be functions of the space-time position, in which case $m$ 
becomes  a scalar potential and  $m'$ a pseudo scalar potential.  We saw above that for a constant mass $m$
the action (\ref{R31})  can be restricted   to the domain ${\cal S}$, preserving   the invariance
under shifts of the Rindler time $\phi$.  
It is clear that if $m$ depends  in $\rho$, but not in $\phi$, the action remains
invariant under translations of $\phi$,  and that the   Hamiltonian is equal to  (\ref{R43}), with $m$
replaced by $m(\rho)$. The same happens  with  the term $\bar{\psi} \gamma^5 \psi$
if $m' = m'(\rho)$. Concerning the vector potential,  the action (\ref{R311}) is invariant provided 

\beq
A_0 \pm A_1 = a_\pm(\rho) \, e^{ \mp \phi},
\label{R312}
\eeq
in which case (\ref{R311})  becomes

\barray 
S & = & \frac{i}{2}
\int_{- \infty}^\infty d \phi \, \int_{\ell_1}^{\infty} d \,   \rho \left[ \chi^\dagger_- (  \partial_\phi + \rho \partial_\rho  + \frac{1}{2}  + i a_+ \rho)  \chi_-    
  \right. \label{R313}  \\
& + &   \chi^\dagger_+ ( \partial_\phi -  \rho \partial_\rho  -   \frac{1}{2}  + i a_- \rho)  \chi_+ \left. +  (i m + m')  \rho  \chi^\dagger_-  \chi_+ +   (i m - m') \rho    \chi^\dagger_+ \chi_-   
\right] . \nonumber 
\earray 
Using gauge transformations one can reduce the number of fields in (\ref{R313}). This issue
will be consider below in a {\em discrete}  realization of this model. 
%To study the model in ${\cal R}_+$, we simply take the limit $\ell_1 \rightarrow 0$.
%The action in ${\cal R}_+$  is gauge invariant  under  local phase transformations of the fields $\chi_\pm$, that
%are compensated by changes in $a_\pm, m, m'$ 
%\barray 
%\chi_\pm & \rightarrow &  e^{ i \alpha_\pm(\rho) } \, \chi_\pm, \label{R314} \\
%a_\pm & \rightarrow  & a_\pm \mp \partial_\rho \alpha_\mp, \nonumber \\
%m + i m' &  \rightarrow &  e^{- i (\alpha_+ - \alpha_-)}  (m + i m' ). \nonumber 
%\earray 
%The gauge group is $U(1) \otimes U(1)$, which  implies that one can gauge away
%two of the four functions $a_\pm, m, m'$,  setting for example, $a_\pm =0$,  or $m'=0$ and $a_+ - a_-=0$. 
%So in ${\cal R}_+$,  the model depends only on two functions. On the other hand, if   $\ell_1$  is finite (model defined in $\cal S$),  
%the action  (\ref{R313}) implies the boundary condition  (\ref{R49}), that is solved by  (\ref{R50}). 
%In this case  the gauge symmetry (\ref{R314}) reduces to the diagonal part, that is $\alpha_+ = \alpha_-$
%and therefore the number of independent parameters increases by one. 
%This issue will be discussed later on. 
The equations of motion derived from (\ref{R313}) are 

\beq
 (  \partial_\phi  \pm  \rho  \partial_\rho \pm  \frac{1}{2} + i a_\pm \rho ) \chi_\mp  +( i m \pm m')   \rho  \chi_\pm = 0, 
\label{R411}
\eeq
and can be written as the Schroedinger equation (\ref{R42}) with  Hamiltonian 

\barray 
H_R  & = &  
\left( \begin{array}{cc}
- i(  \rho \,   \partial_\rho + \frac{1}{2} )   + a_+ \rho   & (m - i m' )\rho \\ 
(m + i m') \rho &  i ( \rho \,   \partial_\rho + \frac{1}{2} )  +   a_- \rho ) \\
\end{array}
\right)  \label{R315} \\
& = &  \sqrt{\rho} \, \hat{p}_\rho \, \sqrt{\rho} \, \sigma^z + 
\frac{a_+ + a_-}{2} \rho \,  {\bf 1}  +  \frac{a_+ -  a_-}{2}\rho\, \sigma^z + 
m \rho \,  \sigma^x + m'  \rho \,  \sigma^y, 
\nonumber 
\earray 
that  acts on the wave functions that satisfy the boundary condition  (\ref{R50}). 

In summary, we have shown in  this section, that the spectrum of the Rindler Hamiltonian, 
in the domain ${\cal S}$  agrees asymptotically with the
average Riemann zeros, under the identifications (\ref{R52c}). However,  there is no trace
of their fluctuations that  depend  on the zeta function on the critical line. This observation is no surprising since
the prime numbers  should be  included  into  the model.  In the next section we shall take a first step in that direction.

\section{Moving mirrors and prime numbers}

Here  we construct an ideal optical system, in Rindler space-time, 
that allows to distinguish  prime numbers from  composite. 
In the next section we shall give a concrete realization of  this optical system in  the Dirac model. 
The system consists of an infinite array of mirrors, labelled by the integers $n=1, 2, \dots,  \infty$,
that have the following properties: 

\begin{itemize}

\item The first mirror, $n=1$, is perfect,  while  the remaining ones, $n > 1$,  
are one-way mirrors (beam splitters) that reflect and transmit the light rays partially. 
The light rays can be replaced by massless fermions. 

\item The mirrors move in Minkowski space-time with uniform accelerations $a_n$,
with $a_1 > a_2 > \dots$, such that $\lim_{n \rightarrow \infty} a_n  =0$. 

\item At time $x^0=0$, the mirrors are placed at the positions $\ell_n = 1/a_n$ (units $c=1$). 

\item The worldlines of the mirrors are contained in  the domain ${\cal S}$,
whose boundary corresponds to the first mirror,  $n=1$, such that  $\rho=\ell_1$. 

\item An observer carries  the first mirror, 
and  sends and receives light rays whose departure and arrival times  she  measures  with  a clock. 

\item The lengths $\ell_n$ are given by

\beq
\ell_n = \ell_1 \, n^{1/2}, \qquad n =1, 2, \dots,  \infty . 
\label{m0}
\eeq
%This choice can be replaced by  $n^{\alpha} \; (\alpha >0$), but the parameter $\alpha$
%can be set to $1/2$ by scaling the clock's  ticks. 

\end{itemize}

\begin{figure}[t]
\begin{center}
{\includegraphics[height= 6.0 cm]{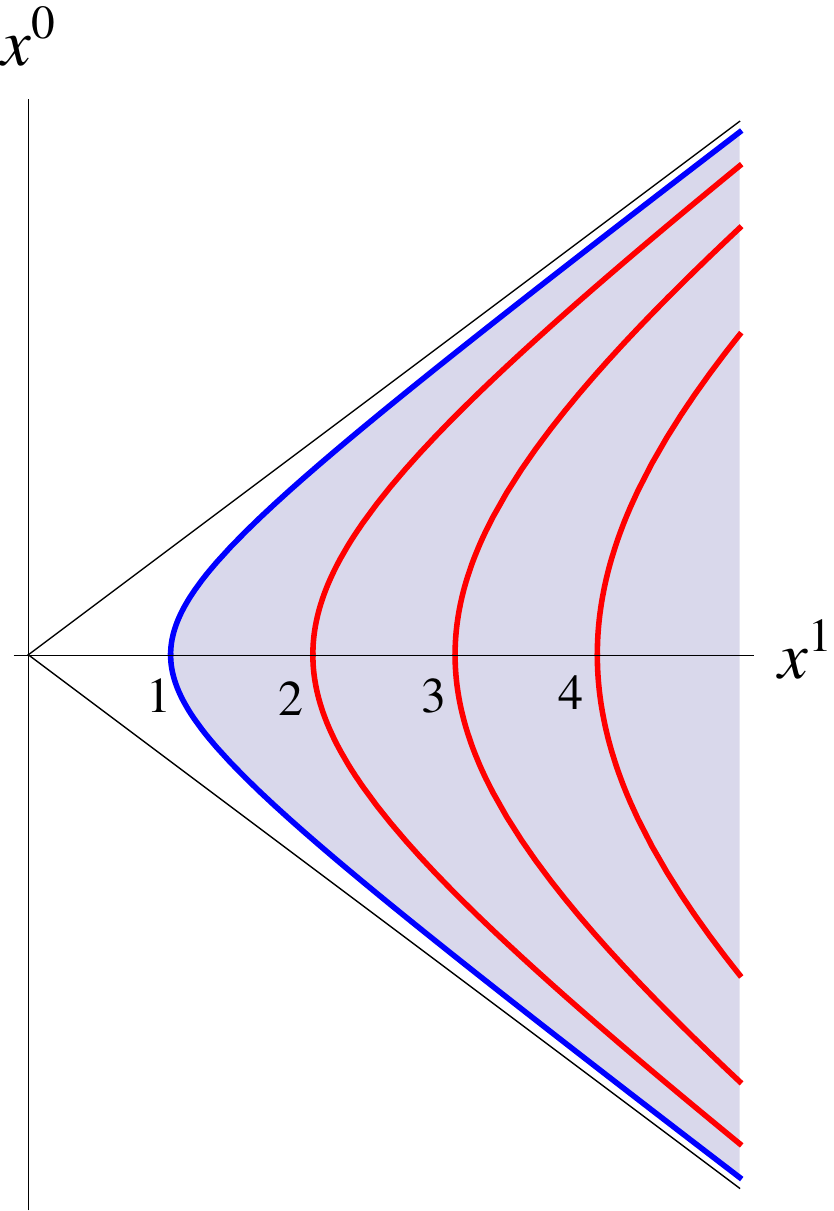}
\hspace{1.5cm}
\includegraphics[height= 5.5 cm]{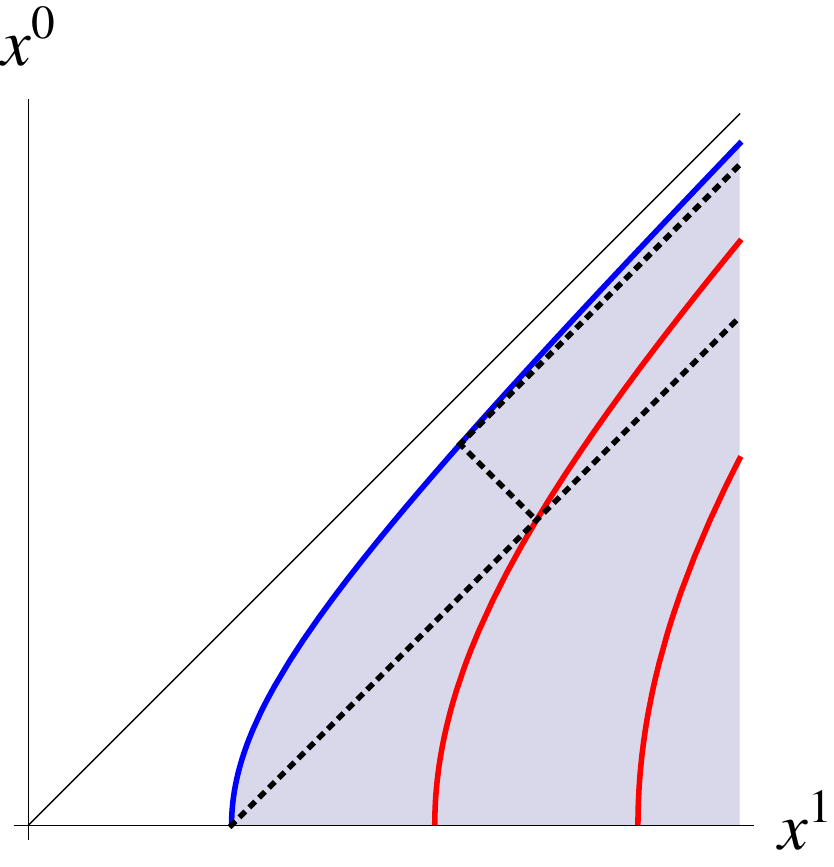}}
\end{center}
\caption{Left: worldlines of the mirrors with accelerations $a_n= 1/n$ that corresponds to the choice
$\ell_n = n \; (n=1,2, \dots )$. The case of eq.(\ref{m0}), is a simple rescaling of lengths.  
${\cal S}$ is the grey  region. Right:  A light ray (dotted line) is sent  at $x^0=0$ by the observer at $x^1 = \ell_1$. 
The ray either passes through  the $2^{\rm nd}$ mirror or is reflected back to the  $1^{\rm st}$ mirror, where is reflected totally. 
} 
\label{mirrors}
\end{figure}

This ansatz can be replaced by  $n^{\alpha} \; (\alpha >0$), but the parameter $\alpha$
can be set to $1/2$ by scaling the clock's  ticks. 
Figure \ref{mirrors} depicts the  mirror's  worldlines satisfying these conditions with $\alpha =1$.

We shall next  study the propagation of light rays in this optical array using 
the laws of special relativity. 
Let us consider a light ray  emanating  from  the point $(\rho_1, \phi_1)$
and reaching  the point $(\rho_2, \phi_2)$,  where $(\rho_i, \phi_i)$ are  the Rindler coordinates. 
Along this trajectory the line element (\ref{R13}) vanishes, 
\beq
d \rho = \pm \rho \,  d \phi 
\rightarrow   \rho_2=  \rho_1 e^{ \pm ( \phi_2 - \phi_1)} \rightarrow \phi_2 - \phi_1 = \arrowvert \log \frac{ \rho_2}{\rho_1} | , 
\label{m1}
\eeq
corresponding to right  moving ($\rho_2 > \rho_1$) or   left moving ($\rho_1 > \rho_2$) rays.
Suppose that the   ray is emitted at the first   mirror at time $\phi_{in}=0$, i.e. 
$(\rho_{\rm in}, \phi_{\rm in}) = (\ell_1, 0)$,  
moves rightwards, reflects on the n$^{\rm th}$ mirror and returns to  the first mirror, 
$(\rho_{\rm end}, \phi_{\rm end}) = (\ell_1, \phi_{n})$.
The value of $\phi_{n}$ follows from eq.(\ref{m1}) and (\ref{m0})
\beq
\phi_{n} = 2  \log \frac{ \ell_n}{\ell_1} = \log n, \qquad n >1. 
\label{m2}
\eeq
which is twice the change in $\phi$ from $\rho= \ell_1$ to $\rho= \ell_n$. 
$\phi_n$ can be measured by  the  observer's clock   traveling with the perfect mirror  where the ray was emitted and received. 
The change in  the clock's proper time is given by $\tau_n^{(1)} = \ell_1 \phi_n$ (see eq.(\ref{R15})), which  in units of $\ell_1 =1$, reads
\beq
\tau_n =  \log n, \qquad n >1. 
\label{m3}
\eeq
Hence,   measuring $\tau_n$,  the observer can find the value of $n$. 
If $n =2$, the ray travels forth and back between the mirrors $n=1$ and $n=2$.
However if $n > 2$, the ray must pass through  the intermediate mirrors
$2, \dots, n-1$. Let us next analyze the  case of two reflections.
Now  the ray is emitted at 
$(\rho_{\rm in}, \phi_{\rm in}) = (\ell_1, 0)$, reaches the mirror $n_1$, returns
to the perfect mirror, reflects again, reaches the mirror $n_2$ and comes back  finally
to the perfect mirror where the clock records  the proper  time, $\tau_{n_1, n_2}$
that  is given by the sum of the intermediate times (\ref{m3}) 
\beq
\tau_{n_1, n_2} = \tau_{n_1} + \tau_{n_2}  = \log (n_1 n_2), \qquad n_{1,2} > 1. 
\label{m4}
\eeq
Hence the measurement of $\tau_{n_1, n_2}$ allows the observer to 
compute the product $n_1 n_2$.  There are   more complicated cases 
as the one illustrated by the following sequence 
\barray 
& 1 \rightarrow n_1 \rightarrow n_2 \rightarrow n_3 \rightarrow 1,  &\label{m5} \\
& 1 < n_1 > n_2 < n_3 > 1,  &
\nonumber 
\earray 
where the  ray emitted by the observer is reflected by the  mirror $n_1$, 
 back  to the mirror $n_2 < n_1$, 
that  reflects the ray forwards to  the mirror $n_3 > n_2$, that reflects the ray
back to the first mirror. The proper time recorded by the clock is 
\beq
\tau_{n_1, n_2, n_3}   =  
 \tau_{n_1} -  \tau_{n_2} + \tau_{n_3}   = \log \frac{n_1 n_3}{n_2}. 
\label{m6}
\eeq
Notice that the case $n_2=1$ reproduces eq.(\ref{m4}). 
It is easy to derive  a general formula for the proper time elapsed  
for  a  trajectory involving  $2k$
intervals,  that starts and ends at the first mirror, 
\barray 
1 \rightarrow n_1 \rightarrow n_2   \rightarrow  &  \dots &  \rightarrow   n_{2 k-1} \rightarrow n_{2k}=1, \label{m7} \\
1 < n_1 > n_2 < &  \dots &   < n_{2 k-1} > n_{2k}= 1,   
\nonumber 
\earray 
and  is  given by 

\beq
\tau_{ \left\{ n_i \right\}}  =  
 \log \frac{ {n_1} {n_3} \dots  {n_{ 2 k -1}} }{{n_2} {n_4} \dots {n_{2k}}}.
\label{m8}
\eeq
The numerator of (\ref{m8}) corresponds to 
reflections:   right  mover $\rightarrow$  left  mover, while the denominator
corresponds to reflections: left   mover $\rightarrow$  right mover. 
The argument of $\log$ in (\ref{m8}) will be  in general a rational number. 
Let us suppose it  is the prime $p$. It is clear,  from (\ref{m3}),
that after one reflection, i.e. $k=1$, the observer 
will detect one ray arriving at   $\tau_p =  \log p$.  Suppose now that the argument of the log 
 is a composite number $n$, say  $4$. At time $\tau_4 =  \log 4$,
the observer  will detect the  ray reflected  from the forth  mirror, but also one ray from 
 two  reflections on the second mirror $\tau_{2,2} = 2 \tau_{2}$, which is of course
the same as $\tau_4$. 
In a real experiment
the two rays arriving at the observer will interfere. The study of this interference
is left to the next section. 

The previous example suggests that
prime numbers correspond to unique paths characterized by observer proper  times
equal to $\log p$. Let us prove this statement in   the case  $k=2$. 
Equation (\ref{m6}) becomes 
\beq
\frac{n_1 n_3}{ n_2} = p, 
\label{m11}
\eeq
where $n_{1,2,3}$ satisfy the constraints  (\ref{m5}). 
According to (\ref{m11}),    $p$ divides the product $n_1 n_3$, so it is a  prime factor
of $n_1$ or $n_3$. In the former case one has
\beq
n_1 = p \,  n'_1 \; \; (n'_1 \geq 1) \rightarrow 
\frac{n'_1 n_3}{ n_2} = 1,
\label{m12}
\eeq
which is a contradiction  because $n_2 < n_3$ by eq.(\ref{m5}). 
The same result holds  if $p$ divides $n_3$. 
The  generalization  to any $k \geq 2$ goes as follows. 
Suppose that  $\tau_{ \left\{ n_i \right\}}   = \log p$, then 
from  eq.(\ref{m8})
\beq
 \frac{ {n_1} {n_3} \dots  {n_{ 2 k -1}} }{{n_2} {n_4} \dots {n_{2k-2}}} = p . 
\label{m13}
\eeq
If $p$ divides, say $n_1$, one gets
\beq
n_1 = p \,  n'_1 \;(n'_1 \geq 1) \rightarrow 
 \frac{ {n'_1} {n_3} \dots  {n_{ 2 k -1}} }{{n_2} {n_4} \dots {n_{2k-2}}} = 1 , 
\label{m14}
\eeq
which cannot be satisfied  because $n_2 < n_3, \dots, n_{2 k -2} < n_{2 k -1}$
by the conditions (\ref{m7}).  This proves that the observer detects a
single ray only when it comes from the reflection on  a {\em prime}  mirror,
while it detects more than one ray when the rays  comes from {\em composite} mirrors. 
This interpretation is purely classical because it presuposes 
 that the rays can be distinguished, and disregards
the interference effects. Both effects have of course to be taken into account 
in a realistic implementation using identical particles,  such as  photons or masless
fermions. The interference pattern emerging for fermions  in this array  will be purpose of the next section. 
In any case, one can easily show that this  classical model can be used to 
 implement the classical  Eratosthenes's sieve to construct prime numbers.

\vspace{0.3 cm}

{\bf Comments:}

\begin{itemize}

\item  We  mentioned  in the introduction, that similarities between counting  formulas in 
Number Theory and  Quantum Chaos led Berry to conjecture 
 the existence of a classical chaotic Hamiltonian  whose 
primitive periodic orbits are labelled by the  primes $p$, 
with periods  $\log   p$, and whose quantization will give  the Riemann zeros as energy levels. 
A classical Hamiltonian with this property has not yet been constructed,
but the mirror  system  presented above, displays some of its properties. 
In particular, the rays  associated to prime numbers behave as primitive 
orbits with a period $\log \, p$. Moreover,  the trajectories
and periods of these {\em primitive} rays are independent of their energy, that is the 
frequency of the light. 

\item One can construct an array of moving mirrors in the domain ${\cal S}_c$ (recall (\ref{R14})) 
with the same properties as the array in ${\cal S}$. The parameters $\ell_n$ that characterize
the array can  be labeled with the negative integers, 
\beq
\ell_n = \ell_1 |n|^{ - 1/2}, \qquad n = -1, - 2,  \dots - \infty, 
\label{m15}
\eeq
where $n= - 1$ is the perfect mirror located at the boundary $\partial {\cal S}_c = \partial {\cal S}$. 
The proper time elapsed  is now given by  $\tau_n = \log | n|$, and  is identical  to  (\ref{m3}) . The relation between the arrays
in ${\cal S}$ and ${\cal S}_c$ corresponds to the inversion transformation  $\rho \rightarrow \ell_1^2/\rho$.

\item The array of mirrors defined  in ${\cal S}$ is   an analogue computer for the multiplication operation or rather
the addition of log's. To implement  the addition operation we shall define   an array of mirror with positions, i.e.  inverse accelerations, 
\beq
\ell_n  = \ell_0 \, e^{ n/2}, \qquad n = 0, 1, 2,  \dots \infty.
\label{m16}
\eeq
Where $n=0$ labels the perfect mirror.  The analogue of eqs.(\ref{m3}) and (\ref{m4}) is 
(with $\ell_0 =1$) 
\barray 
\tau_n   & =   & n , \label{m17} \\
\tau_{n_1, n_2} & = & n_1 + n_2. 
\nonumber 
\earray
This is  an {\em harmonic}  array in the sense  that  the proper times (\ref{m17}),
as well as the discrete spectrum (see Appendix A)   are  given by integer numbers. 
%In section ?? we shall consider quantum versions of these models. 
The {\em harmonic}  array  shows that it is not enough to have accelerated mirrors to generate chaos.
The location/accelerations of the mirrors is essential. In the {\em harmonic}  case, the exponential
separation of the mirrors (\ref{m16}) compensates  the exponential time dependence
of the ray  trajectories (\ref{m1}), that give  rise to a  tessellation of  Rindler space-time. 
In the case of (\ref{m0}), the whole  set of ray trajectories  does not  tessellate the Rindler spacetime 
which is a manifestation of  the arithmetic chaos. 

\end{itemize}

\section{Massless Dirac  fermion with delta function potentials}

In this section we  present  a mathematical  realization of  the moving mirrors  in the Dirac  theory. 
We  start from the massless Dirac action and represent the mirrors by delta
function potentials that are obtained discretizing   the interacting terms
in the general action (\ref{R313}). The mass term becomes a contact
interaction that turns left moving fermions into right moving ones, and viceversa. 
The new  action remains invariant under translations
in  Rindler time, so that  the Rindler Hamiltonian is a conserved quantity 
and we look for its  spectrum using transfer matrix methods. 
The Dirac equation with delta function  potentials 
requires a special  treatment that we describe in detail (see  \cite{SM81}-\cite{CNP97} for the
Dirac equation with delta interactions  in Minkowski coordinates).

\subsection{Discretization  in Rindler variables}

Let us consider  the integral
\beq
\int_{\ell_1}^\infty  d\rho \, \rho \, f(\rho) = \frac{1}{2} \int_{u_1}^\infty du   \, f(u), \quad u = \rho^2, u_1 = \ell_1^2,
\label{dm1}
\eeq
where $f(\rho)$ is a generic function,  and $d \rho \, \rho$
is the radial part of the measure $d^2 x =  d \phi \, d \rho \, \rho$. 
The Rindler time $\phi$  can be easily   incorporated into the  equations. 
We shall  discretize (\ref{dm1}), using the positions  $\ell_n$ of the mirrors
given in eq.(\ref{m0}), which amounts to a partition of  the  half-line $u \geq  u_1$ 
into segments  of equal width $\ell_1^2$,  separated by the points
\beq
u_k = \ell_k^2 =  \ell_1^2  \, k, \quad k=1,2,  \dots.  
\label{dm2}
\eeq
Let us discretize    (\ref{dm1})   as 
\beq
\frac{1}{2} \int_{u_1}^\infty  du   \, f(u) \rightarrow \frac{ \ell_1^2}{2} \sum_{k=2}^\infty f(u_k) \rightarrow 
\frac{ \ell_1^2}{2}  
\int_{u_1}^\infty  du   \, f(u) \sum_{k\geq 2} \delta(u - u_k). 
\label{dm3}
\eeq
If $f$ is a continuous function the last expression of (\ref{dm3})  coincides with the middle one, but if 
$f$  is discontinuous one gets 
\beq
\frac{1}{2} \int_{u_1}^\infty du   \, f(u) \rightarrow 
\frac{ \ell_1^2}{2} \sum_{k=2}^\infty \frac{f(u_k^+) + f(u_k^-)}{2}, 
\label{dm5}
\eeq
where we used 
\barray
\int_{v - \epsilon}^{v + \epsilon} du \, f(u) \delta(u-v) & = & \frac{1}{2} ( f(v^+) + f(v^-)), 
\label{dm6} \\
f(v^\pm)=   \lim_{\epsilon >0 \rightarrow 0} f(v \pm \epsilon).  & &   \nonumber 
\earray 
%
%Eq.(\ref{dm5}) of course  reduces to  (\ref{dm4}) if $f$ is continuous. 
In the limit $\ell_1 \rightarrow 0$, the last expression in eq.(\ref{dm3})  converges towards  the integral (\ref{dm1})
for well behaved  functions. 
Replacing (\ref{dm2}) into  (\ref{dm3})  yields 

\beq
\int_{\ell_1}^\infty d\rho \, \rho   \, f(\rho) \sum_{k\geq 2}  \frac{ \ell_1^2}{ 2\ell_k}  \delta(\rho - \ell_k). 
\label{dm9}
\eeq
One  could use this formula to discretize the mass term in the Dirac action (\ref{R31}). 
Taking  $f = m \bar{\psi} \psi$ this  would yield 

\beq
\frac{i}{2} 
\int_{\cal S}  d \phi \, d\rho \, \rho   \, \bar{\psi} \psi  \sum_{k\geq 2}  \frac{m  \ell_1^2}{ 2\ell_k}  \delta(\rho - \ell_k).
\label{dm10}
\eeq
However, the corresponding Dirac  equation is problematic, because the matching conditions are not
consistent with the equations of motion, 
as shown  in references  \cite{SM81}-\cite{CNP97}. This problem is solved 
by  replacing   the local delta interactions by  separable delta function potentials, that amounts  to  a point splitting. 
More concretely, the integral 

\beq
\int_{\ell_1}^\infty  d\rho \, \rho \, f(\rho) g(\rho)  = \frac{1}{2} \int_{u_1}^\infty du   \, f(u) \,  g(u), 
\label{dm11}
\eeq
should be  replaced by

\barray 
& \frac{ \ell_1^2}{2}  
\int_{u_1}^\infty du \int_{u_1}^\infty   du'   \, f(u) g(u') 
  \sum_{k\geq 2} \delta(u - u_k) \delta(u' - u_k) , 
& \label{dm12} \\
& = \int_{\ell_1}^\infty  d \rho \rho \int_{\ell_1}^\infty d\rho' \rho' f(\rho) g (\rho')  
\sum_{k\geq 2}  \frac{ \ell_1^2}{2 \ell_k^2}  \delta(\rho - \ell_k) \delta(\rho' - \ell_k). 
&  \nonumber 
\earray
If $f$ and $g$ are continuous functions at $\ell_k$, this expression is equivalent to (\ref{dm9}).

\subsection{The Dirac action with delta function potentials} 

Applying  the discretization formula (\ref{dm12}) to the general action (\ref{R313}) one obtains

\beq 
S_{DM} = S_{\rm 0} + S_{\rm int }
\label{dm13}
\eeq
where $S_{\rm 0}$ is the massless action 
%
%\beq
%S_0  = \frac{i}{2} \int_{- \infty}^\infty d \phi \int_{\ell_1}^{\infty} d \rho \, \rho  \, \bar{\psi} 
%\, {\slashed \partial} \,    \psi
%\label{dm14}
%\eeq
\barray
S_0  & = & \frac{i}{2}
\int_{- \infty}^\infty d \phi \, \int_{\ell_1}^{\infty} d \,   \rho \left[ \chi^\dagger_- (  \partial_\phi + \rho \partial_\rho  + \frac{1}{2} )  \chi_-    
+  \chi^\dagger_+ ( \partial_\phi -  \rho \partial_\rho  -   \frac{1}{2} )  \chi_+  \right] ,  \label{dm14} 
\earray 
and $S_{\rm int}$ is the discretization of the mass terms  and vector potential of (\ref{R313}) in the gauge $a_+=a_-$

\barray 
 S_{\rm int} & = & 
 \frac{i}{2}
\int_{\cal S}  d \phi \,  d \rho \,  \rho \, d \rho'  \rho'  \, \sum_{k \geq  2}^{}   \frac{2}{ \ell_k}    \delta(\rho - \ell_k) \delta(\rho' - \ell_k) 
\label{dm15} \\
& \times & 
\left[  i r''_k   \,   \chi_-^\dagger(\rho, \phi) \, \chi_-(\rho', \phi) + 
i r''_k \,   \chi_+^\dagger(\rho, \phi) \, \chi_+(\rho', \phi) \right.  \nonumber  \\
&  & \left. +
  ( i r_k + r'_k)   \chi_-^\dagger(\rho, \phi) \, \chi_+(\rho', \phi) + 
( i r_k - r'_k)   \chi_+^\dagger(\rho, \phi) \, \chi_-(\rho', \phi) \right] , 
\nonumber 
\earray 
where
\beq
r_k =  \frac{ m(\ell_k)  \ell_1^2}{4  \ell_k}, \qquad r'_k =  \frac{ m'(\ell_k)  \ell_1^2}{4  \ell_k},  \qquad 
r''_k  =  \frac{ a_\pm(\ell_k)  \ell_1^2}{4  \ell_k},  \qquad k \geq 2, 
\label{dm15b}
\eeq
are dimensionless parameters that  are equal to the values
taken by the functions $m,m', a_\pm$ in (\ref{R313}) multiplied by $\ell_1^2/(4 \ell_k)$. 
 %The action (\ref{dm13}) is invariant under shifts of $\phi$, that are generated by the Rindler Hamiltonian
%whose spectrum we shall analyze below. 
%If  $m_k = m , \forall k$, the eq. (\ref{dm15})  coincides  with the discretization
%of the mass term in  the Dirac action,  given by  eq.(\ref{dm12}), where 
%$f$ and $g$ are replaced   by $\bar{\psi}$ and $\psi$, respectively. 
One can verify that the (\ref{dm13}) is invariant under the scale 
 transformation 

\beq
\rho \rightarrow  \lambda \rho,  \quad  \chi_\pm \rightarrow \lambda^{-1/2} \,   \chi_\pm,  
 \quad \ell_k \rightarrow \lambda \ell_k, 
\label{dm15c}
\eeq 
so that the physical observables only depend on the parameters $r_k, r'_k, r''_k$. 
 %To simplify notations,  we shall often use the quantity
 % \beq
 %{\slashed h} \equiv  \frac{m \ell_1}{\hbar} 
 %\label{dm15d}
 %\eeq
 %that is dimensionless and equal to $2 \pi$ in the BK, $x/p$ and ${\slashed p}$ models.  
 In the Dirac model considered in the previous section, the mass is constant, which after discretization implies 
 that   $r_n =  m \ell_1/(4 n^{1/2})$.   Later on we shall generalize this eq. to $r_n \propto n^{- \sigma}$, and make contact with the Riemann zeta
 function $\zeta(\sigma + i t)$. 
 
% $m_k = m$, and eqs.(\ref{m0}) and (\ref{dm15b}), it follows that 
 %$r_k \propto 1/k^{1/2}$. To generalize this model, we shall consider the choice
 %
% \beq
%r_k =  \frac{ {\slashed h}}{4  \,  k^\sigma}, \qquad k \geq 2, 
%\label{dm15e}
%\eeq
% where $\sigma$ is a real parameter, which together with ${\slashed h}$  and 
% the parameter $\vartheta$ of the self adjoint extension of the Hamiltonian
% will determine the spectrum. 

\subsection{The Hamiltonian} 

The equations of motion that follows   from (\ref{dm13})  are

\beq 
(  \partial_\phi \pm  \rho \partial_\rho  \pm  \frac{1}{2} )  \chi_\mp 
 + \sum_{k  \geq 2}    \ell_k    \delta(\rho - \ell_k)   \left[  i r''_k ( \chi_\mp(\ell_k^+) + \chi_\mp(\ell_k^-) )   +   (i r_k  \pm r'_k)  \, 
( \chi_\pm(\ell_k^+) + \chi_\pm(\ell_k^-) )   \right]  = 0, \label{dm16}
\eeq
where (\ref{dm6}) has been  used  to integrate around  $\rho = \ell_k$ (to simplify the notation the variable  $\phi$ has been suppressed).  
This equation implies  that  the left and right moving modes 
propagate freely and independently between the positions $\ell_n$, 
\barray
& (  \partial_\phi \pm  \rho \partial_\rho  \pm  \frac{1}{2} )  \chi_\mp  = 0, \quad \rho \neq \ell_k.   & \label{dm19}  
\earray 
Hence  the Hamiltonian is  (recall  (\ref{R43})), 
\beq
H_R = \left( 
\begin{array}{cc}
\sqrt{ \rho} \,  \hat{p}_\rho   \sqrt{\rho} & 0 \\ 
0 & - \sqrt{ \rho} \,  \hat{p}  \sqrt{\rho} \\
\end{array}
\right), \quad \hat{p}_\rho = - i \partial_\rho, \quad  \rho \neq \ell_{n} \; , \forall n. 
\label{dm17}
\eeq
The delta function terms   yield the  
 matching conditions between the wave functions on both sides
 of  $\ell_k$, 
\beq
( \pm 1 + i r''_k)    ( \chi_\mp( \ell^+_k) -  \chi_\mp( \ell^-_k))  
+ (i r_k \pm r'_k)   ( \chi_\pm(\ell_k^+) + \chi_\pm(\ell_k^-) )  = 0, 
 \quad  k  \geq  2, 
\label{dm20} 
\eeq
that can be  written  in matrix form as
\beq
N_{k, +} \, \chi(\ell_k^+)  =  N_{k, -} \, \chi(\ell_k^-) , \qquad k \geq 2 , 
\label{dm201}
\eeq
where $\chi$ is the two component vector (\ref{R42}) and 
\beq
{N}_{k, \pm} = 
\left( \begin{array}{cc}
1 \pm i r''_k  & \pm   ( i r_k + r'_k)  \\
\pm (- i r_k + r'_k) & 1 \mp i r''_k   \\
\end{array}
\right). 
\label{dm202}
\eeq
These matrices are invertible provided 
\beq
\det \, N_{k, \pm} = 1 - r_k^2 - {r'}_k^2  + {r''}_k^2  \neq 0. 
\label{2020}
\eeq
Otherwise,  eqs.(\ref{dm20})  become  a  decoupled   set of equations, hence from  hereafter  we shall assume that
 condition  (\ref{2020})  is satisfied  and leads to 
\beq
 \chi(\ell_k^-)  =  L_{k} \, \chi(\ell_k^+),  \qquad k \geq 2, 
\label{2021}
\eeq
where
\barray
{L}_{k}  & = &  {N}_{k, -}^{-1}  {N}_{k,+}  \label{dm202} \\
& = & 
 \frac{1}{1 - r_k^2 - {r'}_k^2  + {r''}_k^2}  
\left( \begin{array}{cc}
1 + r_k^2 + {r'}_k^2 - {r''}_k^2 + 2 i r''_k     &   2  ( i r_k + r'_k)  \\
2  (- i r_k + r'_k) & 1 + r_k^2 + {r'}_k^2 - {r''}_k^2- 2 i r''_k  \\
\end{array}
\right).  
\nonumber 
\earray 
%
%In addition to eqs.(\ref{dm19}) and (\ref{dm201}), there is a  boundary  conditions at $\rho=\ell_1$ 
Finally, the variation of the action at the boundary $\rho=\ell_1$ gives the condition  (\ref{R50})
\barray 
- i e^{i \vartheta}   \,  \chi_-(\ell_1)  &  = &       \chi_+(\ell_1) \longrightarrow 
\chi(\ell_1)  \propto 
\left( \begin{array}{c}
1 \\
- i e^{i \vartheta} \\
\end{array}
\right), \qquad \vartheta  \in [0, 2  \pi).
\label{dm22} 
\earray

\subsection{Self-adjointness of the Hamiltonian}

One should expect the  Hamiltonian (\ref{dm17}), acting on wave functions subject to 
the BC's (\ref{2021}) and (\ref{dm22}),  to be  self-adjoint. We shall next show that this is indeed
the case.  The scalar product, given by eq. (\ref{R431}), will be written as 
\beq
\langle \chi_1 | \chi_2 \rangle = \sum_{n=1}^\infty  \int_{\ell_n}^{\ell_{n+1}}   d \rho 
\left(  \chi^*_{1,-} \chi_{2,-}   +  \chi^*_{1,+} \chi_{2,+}   \right).
\label{dm281}
\eeq
To show that $H_R$ is a   self-adjoint operator we  follow the approach of
Asorey et  al.  based on the consideration of the boundary conditions that  turns out to be  equivalent to the von Neumann
theorem \cite{AIM05}. The starting point  is the bilinear 
\beq
\langle \chi_1 | H_R \chi_2 \rangle - \langle H_R  \chi_1 |  \chi_2 \rangle \equiv i \Sigma(\chi_1, \chi_2), 
\label{dm282}
\eeq
where $|\chi_{1,2} \rangle$ are two  wave functions.  This quantity measures
the net flux or probability flowing across the boundary of the system, which for a unitary time evolution
generated by $e^{ - i \phi H_R}$ must vanish. The self-adjoint extensions of $H_R$
select subspaces of the total Hilbert space where  $\Sigma(\chi_1, \chi_2)=0$. 
In the case of the Hamiltonian (\ref{dm17}) one finds
\barray
 \Sigma(\chi_1, \chi_2)  & = &  \ell_1 \left(  \chi_{1,-}^*  (\ell_1) \;  \chi_{2,-} (\ell_1) -  \chi_{1,+}^* (\ell_1)  \;  \chi_{2,+} (\ell_1) \right)
\label{dm283} \\
& + & \sum_{n \geq 2 } \ell_{n}  \left[   \chi_{1,-}^*  (\ell_n^+)  \; \chi_{2,-} (\ell_n^+) -  \chi_{1,+}^*  (\ell_n^+)  \;  \chi_{2,+} (\ell_n^+) \right.
\nonumber \\
& & \left. -  \chi_{1,-}^*  (\ell_n^-)  \; \chi_{2,-} (\ell_n^-) +  \chi_{1,+}^*  (\ell_n^-)  \;  \chi_{2,+} (\ell_n^- ) \right]. 
\nonumber 
\earray 
The term proportional to $\ell_1$ already cancels out  by eq.(\ref{dm22}). Imposing the
independent cancellation of the terms proportional to  $\ell_n  \; (n \geq 2)$ yields 
\beq
  \chi_{1,-}^*  (\ell_n^+)  \; \chi_{2,-} (\ell_n^+) -  \chi_{1,+}^*  (\ell_n^+)  \;  \chi_{2,+} (\ell_n^+)
  =  \chi_{1,-}^*  (\ell_n^-)  \; \chi_{2,-} (\ell_n^-) -  \chi_{1,+}^* (\ell_n^-)  \;  \chi_{2,+} (\ell_n^-) 
\label{dm284}
\eeq
which we write as
\beq
\left(   \chi_{1}  (\ell_n^+)| \sigma^z  | \chi_2(\ell_n^+) \right)  = \left(   \chi_{1}  (\ell_n^-)| \sigma^z  | \chi_2(\ell_n^-) \right) , \qquad
\label{dm285}
\eeq
where $| \chi )  = (\chi_-, \chi_+)^t$  and 
$( \chi| = (\chi_-^*, \chi_+^*)$. 
The general solution of (\ref{dm285})   is obtained if $ \chi_{1,2}(\ell_n^-) $ and $\chi_{1,2}(\ell_n^+)$ 
are related by a  transformation 
\beq
 \chi(\ell_n^-, \phi)  =  U_{n} \, \chi(\ell_n^+, \phi), 
\label{dm286}
\eeq
where $U_n$ satisfies 
\beq
U^\dagger_n \, \sigma^z \, U_n = \sigma^z,
\label{dm387}
\eeq
which implies that  $U_n$ belongs to  the  Lie group $U(1) \otimes SU(1,1)$. 
The  non compact character of  $SU(1,1)$ arises in this problem from  to the relative 
minus sign of the $xp$ terms in  the Hamiltonian (\ref{dm17}).
The $U(1)$ factor can be eliminated  by a  phase  transformation  of the field $\chi(\rho)$ in  the interval 
$\rho \in (\ell_n, \ell_{n+1})$, which reduces the group to $SU(1,1)$, that is 
\beq
SU(1,1)  =  \left\{   \left( 
\begin{array}{cc}
a & b \\
b^* & a^* \\
\end{array}
\right),  \; \; a, b \in \Cmath , \; \; |a|^2 - |b|^2 =1 \right\} . 
\label{dm389}
\eeq
  The matrices 
$L_n$ (\ref{dm202}) are of this form,    that ensures that $H_R$ is self-adjoint acting on the wave functions
satisfying the conditions (\ref{2021}) and (\ref{dm22}). 
One can further  reduce the number of parameters in $L_n$  applying 
another $U(1)$   transformation. Let us label the wave functions with an integer $n$ 
\beq
\chi_n(\rho) = \chi(\rho), \qquad \rho \in ( \ell_n, \ell_{n+1}), \quad n =1, 2 , \dots, \infty
\label{dm390}
\eeq
and make  the transformation 
\beq
\chi_{n}(\rho) \rightarrow e^{ i \alpha_n \sigma^z} \, \chi_n(\rho), 
\label{dm391}
\eeq 
that induces the following changes in the boundary values  
\beq
\chi(\ell^+_n) \rightarrow e^{ i \alpha_n \sigma^z} \, \chi(\ell^+_n), \qquad \chi(\ell^-_n) \rightarrow e^{ i \alpha_{n-1} \sigma^z} \, \chi(\ell^-_n), 
\label{dm392}
\eeq 
and in the $L_n$ matrices (recall   (\ref{2021})) 
\beq
L_n \rightarrow e^{ i \alpha_{n-1} \sigma^z} \, L_n \, e^{- i \alpha_n \sigma^z}. 
\label{dm393}
\eeq
The parameters $\alpha_n$ can then be used to bring  (\ref{dm202}) into the form  
\barray
{L}_{n}  & = & 
 \frac{1}{1 - r_n^2 - {r'}_n^2}  
\left( \begin{array}{cc}
1 + r_n^2 + {r'}_n^2     &   2  ( i r_n + r'_n)  \\
2  (- i r_n + r'_n) & 1 + r_n^2 + {r'}_n^2   \\
\end{array}
\right), \quad n \geq 2, 
 \label{Ln}
\earray 
corresponding to an element in the coset $SU(1,1)/U(1)$. We use the same notation  
for the  transformed parameters $r_n, r'_n$ as  in (\ref{dm202}).

In summary, we have shown that the Hamiltonian $H_R$, defined in the domain ${\cal S}$, is self-adjoint acting on wave
functions that satisfy the BC's (\ref{2021}) and (\ref{dm22}); and is  characterized by  
the set   $\{ \ell_n,  r_n, r'_n \}_{n=2}^\infty$ and  $\vartheta$ (by scale invariance (\ref{dm15c})
we set $\ell_1=1$). 
 
%that depend on the choice of the  domain
%\beq  \;  \chi(\ell_k^-)  =  L_{k} \, \chi(\ell_k^+),  \quad k \geq 2, \qquad - i e^{i \vartheta} \chi_-(\ell_1)  = \chi_+(\ell_1).  \label{bc1} 
 %\\ {\cal R}_+  : & &  \;  \chi(\ell_k^-)  =  L_{k} \, \chi(\ell_k^+),  \quad |k|  \geq 2, \qquad \chi(\ell_1^-)  =  L_{1} \, \chi(\ell_1^+)  \label{bc2}
%\eeq 
%where $L_k \; (|k| \geq 2)$ are given by eq.(\ref{Ln}) and $L_1$ by (\ref{395}). 

\subsection{Eigenvalue problem} 

We shall now consider the eigenvalue problem of the Hamiltonian (\ref{dm17}),  
acting on normalizable wave functions  subject to the boundary conditions (\ref{2021}) and (\ref{dm22})
The eigenfunctions of  $H_R$ are given by  

\beq
H_R \, \chi = E \, \chi \rightarrow  - i  \left( \rho \, \partial_\rho + \frac{1}{2} \right )  \chi_\mp = \pm E \, \chi_\mp 
\rightarrow  \chi_\mp \propto \rho^{ - 1/2  \pm i E}.
\label{dm24}
\eeq 
so for the $n^{\rm th}$  interval (recall (\ref{dm390}))

\beq
\chi_{\mp, n}(\rho) = e^{ \pm i \pi/4}  \frac{A_{\mp, n}}{ \rho^{  1/2 \mp  i E}}, \qquad \rho \in (\ell_n, \ell_{n+1}) , 
\label{dm25}
\eeq
where $A_{\mp, n}$ will depend,  in general, on the eigenenergy $E$ and  $\rho$ is  measured
in units of $\ell_1$ (in what follows we  take $\ell_1=1$). The phases $e^{\pm i \pi/4}$ have been introduced 
by   analogy with the eigenfunctions  (\ref{R47}).
The values of $\chi$ on   both sides of $\rho= \ell_n$ are given by 

\barray 
\chi_\mp(\ell^+_n) &  = &  \lim_{\epsilon \rightarrow 0^+} \chi_\mp(\ell_n + \epsilon) = \chi_{\mp, n}(\ell_n) = 
e^{ \pm \frac{ i \pi}{4}}    \frac{A_{\mp, n}}{ \ell_n^{1/2  \mp  i E } }\label{bc3}, \\
\chi_\mp (\ell^-_n) &  = &  \lim_{\epsilon \rightarrow 0^+} \chi_\mp(\ell_n - \epsilon) = 
\chi_{\mp, n-1}(\ell_n) =  e^{ \pm \frac{ i \pi}{4}}   \frac{A_{\mp, n-1}}{ \ell_n^{ 1/2  \mp  i E } }.  \nonumber 
\earray 
%hence 
%
%\barray 
%\chi_\mp  (\ell_n^+)  
%& = &  e^{ \pm \frac{ i \pi}{4}}    \frac{A_{\mp, n}}{ \ell_n^{1/2  \mp  i E } }, \label{dm28} \\
%\chi_\mp (\ell_n^-)  
%& = &  e^{ \pm \frac{ i \pi}{4}}   \frac{A_{\mp, n-1}}{ \ell_n^{ 1/2  \mp  i E } }.   \nonumber 
%\earray 
Let us write these equations in matrix form in terms of   the vector 
\beq
| {\bf 	A}_n   \rangle = 
\left( \begin{array}{c}
A_{-, n} \\
A_{+, n}  \\
\end{array}
\right),  
\label{dm33}
\eeq
and the matrix
\beq
{K}_{n, \pm} = 
\left( \begin{array}{cc}
e^{ i \pi/4} \,  \ell_n^{ i E}  & 0 \\
0 & e^{ - i \pi/4} \,  \ell_n^{ - i E}   \\
\end{array}
\right), 
\label{dm34}
\eeq
such that (\ref{bc3}) read
\barray 
| \chi(\ell^+_n) \rangle  & =  &   \ell_n^{-1/2} \,  K_n \, | {\bf 	A}_n   \rangle, 
\label{dm29} \\
| \chi(\ell^-_n) \rangle  & =  &  \ell_n^{-1/2} \, K_n \, | {\bf 	A}_{n-1}   \rangle. 
\nonumber
\earray 
Plugging these eqs. into the  matching  condition (\ref{2021}) yields  
\beq
K_n \, | {\bf 	A}_{n-1}   \rangle = L_n \, K_n \, | {\bf 	A}_n   \rangle, 
\label{dm30}
\eeq
so 
\beq
 | {\bf 	A}_{n-1}   \rangle = T_n \, | {\bf 	A}_n   \rangle, \qquad T_n = K_n^{-1} \, L_n \, K_n, 
\label{dm31}
\eeq
and substituting (\ref{Ln}) 
\barray
{T}_{n}  & = & 
 \frac{1}{1 - r_n^2 - {r'}_n^2}  
\left( \begin{array}{cc}
1 + r_n^2 + {r'}_n^2     &   2  ( r_n -i  r'_n) \,  \ell_n^{ - 2 i E}   \\
2  (r_n + i  r'_n) \ell_n^{ 2 i E}  & 1 + r_n^2 + {r'}_n^2   \\
\end{array}
\right). 
 \label{dm32}
\earray 
To simplify the notations  let us   define 
\beq
\varrho_n = r_n - i {r'}_n, 
\label{dm33}
\eeq
so that 
\barray
{T}_{n}  & = & T(E, \varrho_n, \ell_n)= 
 \frac{1}{1 - |\varrho_n|^2}  
\left( \begin{array}{cc}
1 + |\varrho_n|^2    &   2  \varrho_n\,  \ell_n^{ - 2 i E}   \\
2  \varrho_n^*\ell_n^{ 2 i E}  & 1 + |\varrho_n|^2    \\
\end{array}
\right).
 \label{dm34}
\earray 
The parameters $\varrho_n$ have  the meaning of reflections amplitudes associated
to the $n^{\rm th}$ mirror. The absence of a mirror at the position $\ell_n$ is expressed  by the condition
$\varrho_n=0$, so that   ${\bf A}_{n-1} = {\bf A}_n$. 
 Concerning the condition (\ref{dm22}), equation   (\ref{dm25}) implies 
\beq
 e^{  i \vartheta} \, A_{-, 1} =  A_{+, 1}.  \label{dm35} 
\eeq
that can be written as  
\beq
 | {\bf 	A}_1   \rangle = | {\bf 	A}_1 (\vartheta)   \rangle  = 
\left( \begin{array}{c}
1 \\
e^{i \vartheta}  \\
\end{array}
\right) . 
\label{dm36}
\eeq
%Similarly the condition (\ref{395}) becomes
%\beq
% | {\bf 	A}_{-1} = T_1 \,  | {\bf A}_1    \rangle, \qquad T_1(\vartheta)  = K_1^{-1} L_1  K_1 = L_1 = e^{ i \vartheta \sigma^z/2}. 
%\label{dm37}
%\eeq
Therefore, the  eigenvalue problem has  been  reduced  to  find the  energies $E$ for which the amplitudes ${\bf A}_n$,  satisfying 
  (\ref{dm31}) and (\ref{dm36}),  yield  wave functions (\ref{dm25}) that  are normalized in the discrete sense
  (Kronecker delta function) or in the continuous sense (Dirac delta function), corresponding to the discrete
  or continuum spectrum of the Hamiltonian.  To this aim,  we shall also need the norm and scalar product of the wave functions written  in terms
of the amplitudes.
%The  norm of  (\ref{dm25}) is given by (see  eq.(\ref{R41}))  
%
%\beq
%\langle \chi_1 | \chi_2 \rangle = \sum_{n=1}^\infty  \int_{\ell_n}^{\ell_{n+1}}    d \rho 
%  \left(  \chi^\dagger_{1,-} \chi_{2,-}   +  \chi^\dagger_{1,+} \chi_{2,+}   \right).
%  \label{dm281}
%  \eeq
Using eq.(\ref{dm281}), the norm of (\ref{dm25})  is given by  
\beq
\langle \chi  | \chi  \rangle = 
\sum_{n=1}^\infty \log \frac{\ell_{n+1}}{ \ell_n}  \, \langle {\bf A}_n | {\bf A}_n \rangle, \quad \qquad  
\langle {\bf A}_n | {\bf A}_n \rangle =     |A_{-,n} |^2 + |A_{+, n}|^2  , 
\label{dm282}
\eeq  
and the scalar product of two eigenfunctions with energies $E_1$ and $E_2$ by 
%
%
%\barray 
%\langle \chi_1 | \chi_2 \rangle  & = & \frac{1}{ i E_{12}}  \sum_{n=1}^\infty 
%\left\{  \left[  \ell_{n+1}^{ i E_{12}} -  \ell_{n}^{ i E_{12}} \right]
%{{A}^{(1)}_{-,n}}^* {A}^{(2)}_{-,n}  
%\right. \label{dm283} \\
%& - & \left.  \left[ \ell_{n+1}^{- i E_{12}} - \ell_{n}^{- i E_{12}} \right]
%{{A}^{(1)}_{+,n}}^* {A}^{(2)}_{+,n}  
%\right\} 
%\nonumber 
%\earray 
%
%
\barray 
\langle \chi_1 | \chi_2 \rangle  & = & \frac{1}{ i E_{12}}  \sum_{n=1}^\infty 
\left\{  \left[  \ell_{n+1}^{ i E_{12}} -  \ell_{n}^{ i E_{12}} \right]
{{A}^{(1)}_{-,n}}^* {A}^{(2)}_{-,n}  - \left[ \ell_{n+1}^{- i E_{12}} - \ell_{n}^{- i E_{12}} \right]
{{A}^{(1)}_{+,n}}^* {A}^{(2)}_{+,n}  
\right\} 
\label{dm283}
\earray 
where $E_{12} = E_{1} - E_{2} \neq 0$ and ${{A}_{\pm, n}}^*$ is the complex conjugate of ${A}_{\pm, n}$. 
Since the  Hamiltonian is self-adjoint this product will  vanish.  

\subsection{Semiclassical approximation}

The recursion relation (\ref{dm31}), together with the initial condition  (\ref{dm36}), 
gives all the vectors ${\bf A}_k$ in terms of  ${\bf A}_1(\vartheta)$
\beq
| {\bf A}_{k}  \rangle   = T_k^{-1} T_{k-1}^{-1} \cdots T_2^{-1} | {\bf A}_{1} (\vartheta) \rangle, \quad k \geq 2.
\label{t0}
\eeq
Except for   some simple cases, as the {\em harmonic}  model (see Appendix A),  it will impossible  to find 
close  analytic expressions of  the product of matrices appearing in (\ref{t0}). The only hope
to made progress is to evaluate (\ref{t0}) in the  limit where the coefficients
$\varrho_n$ are infinitesimally small. We shall  then assume  that $\varrho_n$
are proportional to a parameter $\varepsilon$ that will be taken to zero
at the end of the computation. This parameter plays the role of Planck's 
constant,  so the limit $\varepsilon \rightarrow 0$, will be interpreted as semiclassical. 
This interpretation is supported by the discretization of the massive Dirac equation,
that led to eq.(\ref{dm15b}), according to which $\rho_n = m \ell_1/(4 n^{1/2})$. The connection with the
average Riemann zeros was achieved for   $m \ell_1 = 2 \pi$, that corresponds
in the semiclassical $xp$ Berry-Keating model to the Planck constant $\ell_x \ell_p = 2 \pi \hbar$.

Taking  $\varrho_n  = O(\varepsilon)$,  the matrix $T_n$ given in  eq.(\ref{dm34}) can be replaced  in the limit  
 $\varepsilon \rightarrow 0$, by 
\barray
{T}_{n}  & \approx  &  {\bf 1} + \tau_n \approx e^{ \tau_n} + O(\varepsilon^2),  
\quad \tau_n  = 
\left( \begin{array}{cc}
0   &   2  \varrho_n\,  \ell_n^{ - 2 i E}   \\
2  \varrho_n^*\ell_n^{ 2 i E}  & 0   \\
\end{array}
\right).
 \label{t1}
\earray 
$T_n$ can be expressed exactly  as the exponential of a matrix of the form of $\tau_n$, 
but in the limit $\varepsilon \rightarrow 0$, it  will converge towards the expression given in (\ref{t1}) up to order $\varepsilon^2$. 
Plugging (\ref{t1}) into (\ref{t0}) yields 
\beq
| {\bf A}_{k}  \rangle  =   e^{ - \tau_k}    e^{ - \tau_{k-1}}    \dots  e^{ - \tau_2}   | {\bf A}_{1} (\vartheta) \rangle + O(\varepsilon^2). 
\label{t2}
\eeq
%Note that $e^{\tau_n}$, like  $T_n$,  is a matrix of $SU(1,1)/U(1)$.
It is convenient to define  a matrix $\tau_1$, of order $\varepsilon$,
corresponding to the choice  $n=1$ of $\tau_n$,  which  does not depend on $E$ because  $\ell_1=1$. 
The vector  $| {\bf A}_{1} (\vartheta) \rangle$ can  be replaced by 
$e^{- \tau_1} | {\bf A}_{1} (\vartheta') \rangle$, where $\vartheta'$ is equal to $\vartheta$ up to terms of  order $\varepsilon$. 
Eq.(\ref{t2}) can then be written as 
\beq
| {\bf A}_{k}  \rangle  =   e^{ - \tau_k}    e^{ - \tau_{k-1}}    \dots  e^{ - \tau_1}   | {\bf A}_{1} (\vartheta) \rangle + O(\varepsilon^2). 
\label{t2b}
\eeq
where  $\vartheta'$ has been replaced by  $\vartheta$ since they become the  same quantity in the limit $\varepsilon \rightarrow 0$. 
The product of exponentials of matrices can be approximated by the Baker-Campbell-Haussdorf formula  \cite{GP90} 
\beq
e^{ \varepsilon A} e^{ \varepsilon B} = e^{ \varepsilon (A+B)} + O(\varepsilon^2), 
\label{t3}
\eeq
that yields 
\beq
| {\bf A}_{k}  \rangle    =     | {\bf A}_{k}^T  \rangle + O(\varepsilon^2), \qquad   | {\bf A}_{k}^T  \rangle\equiv e^{ -  \sum_{n=1}^k \tau_k}  | {\bf A}_{1} (\vartheta) \rangle.
\label{t4}
\eeq
Let us next define 
\barray
e^{ - i \Phi_k} \, R_k &  = &  \sum_{n=1}^k  \varrho_n \, \ell_n^{- 2 i E}, \qquad e^{  i \Phi_k} \, R_k  = 
  \sum_{n=1}^k  \varrho_n^* \, \ell_n^{ 2 i E},  \label{t5}  
\earray 
where  $R_k$ and $\Phi_k$ are real for real values of  $E$. The factor 2 multiplying $\varrho_n$  in eq.(\ref{t1}) 
can be absorbed into the parameter $\varepsilon$, so without loss of generality we can write 
\beq
  | {\bf A}_{k}^T  \rangle =  e^{ - R_k  ( \cos \Phi_k \sigma^x + \sin \Phi_k \sigma^y)}  | {\bf A}_{1} (\vartheta) \rangle, 
\label{t6}
\eeq
whose norm is
\beq
\langle {\bf A}_{k}^T   | {\bf A}_{k}^T  \rangle =  e^{ 2 R_k} ( 1 - \cos( \Phi_k - \vartheta) ) +  e^{- 2 R_k} ( 1 + \cos( \Phi_k - \vartheta) ). 
\label{t7}
\eeq
The discrete eigenvalues of the Hamiltonian are those for which the norm (\ref{dm282}) is finite,
which is ensured if (\ref{t7}) vanishes sufficiently fast when $k \rightarrow \infty$. The approximation
we have performed above is a sort of inverse Trotter-Suzuki decomposition, e.g.  $\lim_{n \rightarrow \infty}
(e^{A/n} e^{ B/n})^n = e^{ A+B}$, where  a   product of exponentials of non commuting operators
is replaced by the exponential of their sum  \cite{T59,S76}. Let us consider some examples.

\subsubsection{Harmonic model}

This model is  defined by the parameters (see eq.(\ref{m16}) with $\ell_0 =1$)

\beq
\ell_n = e^{n/2},   \quad \varrho_n = \varepsilon, \quad n =0 , \dots, \infty  . 
\label{p1}
\eeq
Let us remind that in this case  the mirrors (delta functions) are labelled by  $n=1, 2, \dots$, while the boundary
is located   at $\ell_0=1$.  $\varepsilon$ can be  positive or  negative and is the  semiclassical parameter. 
This model has an exact solution  given  in Appendix A, that allow us to  verify the semiclassical approximation
done   below. 
From (\ref{t5}) one finds
\barray
e^{ - i \Phi_k} \, R_k &  = &  \varepsilon  \sum_{n=0}^k  e^{-  i n  E} =
\left\{ \begin{array}{cc} 
\varepsilon (k +1)  & E \in  2 \pi \Zmath \\
 \varepsilon \, \frac{ \sin( (k+1) E/2)}{ \sin (E/2)} e^{- i k E/2}  &   E \notin  2 \pi \Zmath \\
 \end{array}
 \right.  \ . 
 \label{p2}  
\earray 
In  the case $E \in  2 \pi \Zmath$, one  can choose $R_k = \varepsilon (k+1)$ and
$\Phi_k=0$. This identification  is not unique but simplifies the calculation. The norm (\ref{t7}) 
is given by  
\beq
E \in  2 \pi \Zmath \rightarrow 
\langle {\bf A}_{k}^T   | {\bf A}_{k}^T  \rangle =  e^{ 2 \varepsilon (k+1) } ( 1 - \cos  \vartheta ) +  e^{- 2 \varepsilon (k+1) } ( 1 + \cos  \vartheta ), 
\label{p3}
\eeq
while  the norm of the wave function $\chi^T$,  whose amplitudes are  ${\bf A}_{k}^T$, reads    (see eq.  (\ref{dm282}))  
\beq
\langle \chi^T  | \chi^T   \rangle =  \frac{1}{2} 
\sum_{n=0}^\infty \langle {\bf A}_{n}^T   | {\bf A}_{n}^T  \rangle  =  \frac{1}{2} 
\sum_{n=0}^\infty \left[  e^{ 2 \varepsilon( n+1)  } ( 1 - \cos  \vartheta ) +  e^{- 2 \varepsilon (n+1) } ( 1 + \cos  \vartheta ) \right]. 
\label{p4}
\eeq
This series diverges for all values of  $\vartheta$ different from  $0$ and $\pi$. 
This means that the energies  $E = 2 \pi n$, do not belong to the spectrum.
On the other hand,  if 
\beq
\vartheta =0 \; \;  {\rm for } \,  \; \varepsilon > 0,  \quad {\rm or} \quad  \vartheta =\pi \; \;  {\rm for } \, \; \varepsilon < 0
\label{p5}
\eeq
the state has the   norm 
\beq
\langle \chi^T  | \chi^T   \rangle =  \frac{1}{ e^{  2 |\varepsilon |} -1 } \rightarrow  \frac{1}{2 |\varepsilon|}, \quad \varepsilon \rightarrow 0, 
\label{p6}
\eeq
The parameter $\varepsilon$ can be absorbed into the normalization constant of the state. 
Hence, in the two cases   (\ref{p5}) there is  an infinite number of normalized
states with energies $E_m = 2 \pi m \; \;  (m \in \Zmath)$. 

In the case $E \notin  2 \pi \Zmath$, one can choose
\beq
R_k =  \varepsilon \, \frac{ \sin( (k+1)  E/2)}{ \sin (E/2)}, \qquad \Phi_k = \frac{k E}{2}. 
\label{p7}
\eeq
The limit $E \rightarrow 2 \pi m$, yields  $R_k \rightarrow \varepsilon (-1)^{m k} (k+1), e^{ i \Phi_k} \rightarrow (-1)^{m k}$
and one recovers  the previous result.  If   $E \notin  2 \pi \Zmath$,  eq.(\ref{p7}) shows that $|R_k| < | \varepsilon/\sin (E/2)| \; \forall k$,
hence the norm of ${\bf A}^T_k$ will be bounded and the corresponding states will be normalizable
in terms of  Dirac delta functions, e.g.  they belong to the continuum spectrum. 

In summary, the spectrum of Hamiltonian of the {\em harmonic}   model,    in the limit $\varepsilon \rightarrow 0$, is given by
the union of a continuum part, ${\rm Spec}_c$,  and a discrete part,   ${\rm  Spec}_d$,  where
\barray
\vartheta  = 0, \;    \varepsilon   \rightarrow   0^+ \; \;   {\rm or} \;  \;  \vartheta  = \pi  , \;    \varepsilon   \rightarrow  0^-    & : & 
{\rm Spec}_c \rightarrow  \Rmath - 2 \pi \Zmath, \; Å
{\rm Spec}_d  =  2 \pi \Zmath  \label{p71} 
\\
\vartheta  \neq  0 \;   {\rm and }  \;   \pi, \; \;    \varepsilon   \rightarrow   0^\pm   \quad   \qquad    & : & {\rm Spec}_c \rightarrow  \Rmath - 2 \pi \Zmath, \; 
{\rm Spec}_d  =  \emptyset  
\label{p71b}
\earray 
Note that in the case $\vartheta \neq 0, \pi$, the energies $E= 2 \pi n$ are missing in the continuum. 
For finite values of  $\varepsilon$ the spectrum is given by (see Appendix A) 
\barray
\vartheta  = 0, \;    \varepsilon   >   0 \; \;   {\rm or} \;  \;  \vartheta  = \pi  , \;    \varepsilon   < 0    & : & {\rm Spec}_c  = 
  \cup_{n=- \infty}^\infty  2 \pi
[ n + \delta, n + 1 - \delta]  , \; 
{\rm Spec}_d = 2 \pi \Zmath , 
\label{p72} 
\earray 
where  $\sin ( \pi \delta) = 2  \varepsilon/(1 + \varepsilon^2)$ (see eq.(\ref{ap16})), so if 
 $\varepsilon \rightarrow 0$ the gap  between the intervals   $ 2 \pi [ n + \delta, n + 1 - \delta]$ closes
and the  continuum spectrum   becomes $\Rmath - 2 \pi \Zmath$, and the discrete spectrum $2 \pi \Zmath$
as in (\ref{p71}). When  $\vartheta \neq 0, \pi$, one also recovers the spectrum in (\ref{p71b}).

Let us consider another example with the same values of $\ell_n$ but exponential decaying reflection amplitudes
\beq
\ell_n = e^{n/2},   \quad \varrho_n = \varepsilon \, e^{-  \lambda n} \; (\lambda >0), \quad n =0 , \dots, \infty  , 
\label{h1}
\eeq
that yields 
\beq
e^{- i \Phi_k } R_k = \varepsilon \, e^{ - ( \lambda + i E) k/2} \, \frac{ \sinh (( \lambda + i E) (k+1)/2)}{ \sinh (( \lambda + i E)/2)}.
\label{h2}
\eeq
If $\lambda =0$, one recovers eq.(\ref{p2}). Let us  see if the discrete spectrum of the previous model 
survives  when $\lambda >0$. Taking   $E \in 2 \pi \Zmath$  yields 
\beq
E \in 2 \pi \Zmath \rightarrow 
e^{- i \Phi_k } R_k = \varepsilon \, \frac{1 - e^{ - \lambda (k+1)}}{ 1 - e^{- \lambda}}. 
\label{h3}
\eeq
Hence for any finite value of $\lambda$, the quantities $R_k$ are bounded, and therefore  the state belongs to the continuum,
e.g. the discrete spectrum disappears, but it can  be recovered in the limit $\lambda \rightarrow 0$, where
$R_k \rightarrow  \varepsilon (k+1)$.

%so we can choose, $\Phi_k =0$, for $\varepsilon >0$, and
%
%\beq
%R_k = \varepsilon \, e^{ - \lambda (k+1)/2} \, \frac{ \sinh ( \lambda  k/2)}{ \sinh ( \lambda/2)} \stackrel{k \gg 1 }{\longrightarrow} 
% \varepsilon \, \, \frac{ e^{ - \lambda/2} }{ \sinh ( \lambda/2)}  \stackrel{\lambda  \rightarrow  0 }{\longrightarrow}  \frac{ 2 \varepsilon}{\lambda}. 
%\label{h4}
%\eeq
%Hence for any finite value of $\lambda$, the quantities $R_k$ are bounded, so that the state belongs to the continuum,
%but as $\lambda \rightarrow 0$ they diverge,  in agreement with the previous results. 

\subsubsection{Polylogarithm   model} 

This model is defined by the parameters (see eq.(\ref{m0}) with $\ell_1=1$ ) 
\beq
\ell_n = n^{1/2} ,   \quad \varrho_n = \varepsilon  \frac{ e^{ - \lambda n}}{n^\sigma} \; \; (\sigma >0, \; \lambda >0), \quad n =1, \dots, \infty . 
\label{p8}
\eeq
The positivity conditions on  $\sigma$ and $\lambda$ ensures that $\varrho_n \rightarrow 0$ as $n \rightarrow \infty$,
including the case where $\lambda=0$. 
The norm of the wave function $\chi^T$  is given by (see eq.(\ref{dm282})) 
\beq
\langle \chi^T  | \chi^T   \rangle =  \frac{1}{2} 
\sum_{n=1}^\infty  \log \left(1 + \frac{1}{n} \right)  \langle {\bf A}_{n}^T   | {\bf A}_{n}^T  \rangle, 
\label{p9}
\eeq
so  its convergence depends on the asymptotic behavior of $\langle {\bf A}_{n}^T   | {\bf A}_{n}^T  \rangle/n$.
The ansatz  (\ref{p8}) yields 
\barray
e^{ - i \Phi_k} \, R_k &  = &  \varepsilon  \sum_{n=1}^k  \frac{ e^{ - \lambda n}}{n^{\sigma + i E}}, 
 \label{p10}  
\earray 
that in  limit $k \rightarrow \infty$ becomes 
\barray
e^{ - i \Phi_\infty} \, R_\infty  &  = & \lim_{k \rightarrow \infty} e^{ - i \Phi_k} \, R_k=  \varepsilon  Li_{\sigma + i E}(e^{- \lambda}) , 
 \label{p11}  
\earray 
where $Li_s(z)$ is the polylogarithm function \cite{wiki}. 
In the limit $ \lambda \rightarrow  0^+$ there is  the expansion
\barray
e^{ - i \Phi_\infty} \, R_\infty &  = &  \varepsilon  \left[\lambda^{ \sigma-1  + i E}\,   \Gamma( 1 - \sigma - i E)  +  
 \zeta( \sigma + i E)  + O(\lambda)   \right]. 
 \label{p12}  
\earray
If  $\sigma > 1$, the term proportional to $\lambda^{  \sigma -1 + i E}$ drops out and one is left with  
\barray
 \sigma > 1 , \;   
\lambda \rightarrow 0 \rightarrow 
e^{ - i \Phi_\infty } \, R_\infty  
 &  =  &  \varepsilon    \zeta( \sigma + i E) , 
 \label{p13}  
\earray
which  implies  that the  norm of ${\bf A}_k^T$ converges towards a  constant, for any value of $\vartheta$, and therefore  the series (\ref{p9}) 
diverges logarithmically signaling  a continuum of states normalizable
in the Dirac delta sense. 
%
%\noindent
If  $\sigma <  1$, the term proportional to $\lambda^{ \sigma -1 + i E}$ dominates hence 
\barray
\sigma <  1 , \;   
\lambda \rightarrow  0   \rightarrow 
e^{ - i \Phi_\infty} \, R_\infty &  \simeq &  \varepsilon   \lambda^{ \sigma -1 + i E} \;  \Gamma( 1 - \sigma - i E) ,  \label{p14}  
\earray
leading to 
\beq
R_\infty \simeq  |\varepsilon   \lambda^{\sigma-1}  \Gamma( 1- \sigma  - i E)|, \quad e^{ - i \Phi_\infty}  \simeq  {\rm sign} \, (\varepsilon)  
\lambda^{i E}
\left(\frac{ \Gamma( 1 - \sigma - i E)}{  \Gamma( 1 - \sigma + i E)} \right)^{1/2},
 \label{p15}  
\eeq
If $\lambda$ is kept fixed, the norm of ${\bf A}_k^T$ is bounded and one gets again a continuum spectrum. 
Let us see  what happens in the limit  $\lambda \rightarrow 0$. Here 
 $R_\infty \rightarrow + \infty$, and therefore the norm of ${\bf A}_k^T$ does not blow up, and actually 
 converges to zero,   if and only if
\beq
\cos( \Phi_\infty - \vartheta) = 1 \Longrightarrow   {\rm sign} \, (\varepsilon)  \lambda^{-i E}
\left(\frac{ \Gamma( 1 - \sigma + i E)}{  \Gamma( 1 - \sigma - i E)} \right)^{1/2} = e^{i \vartheta},
\label{p16}
\eeq
This  result suggests that a regularized version of this model, involving the limit $\lambda \rightarrow 0^+$, 
 should contain eigenstates satisfying (\ref{p16}). 
For $E>> 1$, the Stirling formula provides the asymptotic behavior 
\barray
n(E) & \simeq & \frac{E}{2 \pi} \log \frac{1}{\lambda} + \frac{ E}{2 \pi} \log \frac{ E}{ 2 \pi e}
+ \frac{1}{4} \left( \frac{3}{2} - \sigma  - {\rm sign} \, \varepsilon \right) - \frac{ \vartheta}{2 \pi}, \;  E \gg 1, 
 \label{p17} 
\earray 
that   in the limit $\lambda \rightarrow 0$,
becomes a continuum. The parameter $\lambda$ regularizes the model and in some respects
is analogue to  the cutoff $\Lambda$ in  Connes's  $xp$ model, with $\Lambda \propto 1/\lambda$. 
However, in eq.(\ref{p17}) the term $E/(2 \pi) \log E/( 2 \pi e)$, appears with an opposite
sign as compared with Connes model,  and it  does not have the meaning of missing spectral lines, 
but rather of  a finite energy  correction.  Notice also that if $\sigma = 1/2$, the spectrum has the symmetry
$E \leftrightarrow - E$, provided $\vartheta =0$ if $\varepsilon >0$ or $\vartheta =\pi$ if $\varepsilon <0$,
like in the harmonic  model analyzed above. 

In summary, for   $\sigma >1$ the spectrum is a continuum  related to the  Riemann zeta function $\zeta(\sigma+ i E)$.
This result  is consistent with the studies carried by several authors in the past 
where   the zeta function appears in connection to the  scattering states of some physical system \cite{PF75}-\cite{J03}. 
However, for  $0 < \sigma < 1$  the
connection with the zeta function is lost and only the smoothed {\em zeros} appear   as a finite
size correction to the level counting formula,   in analogy with Connes's version on the $xp$ model, but with opposite sign. 
We are then forced  to  look for another
ansatz for the reflection coefficients $\varrho_n$, if the {\em zeros} are to be realized as discrete eigenvalues
of the Hamiltonian $H_R$ in the semiclassical limit. A hint is provided by the {\em harmonic} model, where
the eigenvalues  $E_n = 2 \pi n$, arises from the blow up of $R_k(E_n)$ 
as $k \rightarrow \infty$.   This property leads us to the next example.

\subsubsection{Riemann  model} 

The  model is defined by the parameters 
\beq
\ell_n = n^{1/2} ,   \quad \varrho_n = \varepsilon \,  \frac{  \mu(n)}{n^\sigma} \; \; (\sigma >0), \quad n =1, \dots, \infty , 
\label{p18}
\eeq
where $\mu(n)$ is the Moebius function that vanishes  if $n$ contains as divisor  the square of a prime,  and it is equal to 1 ($-1)$
if $n$  is the product of an even (odd) number of distinct  primes, that is \cite{Apostol}
\beq
\mu(n) = \left\{
\begin{array}{cl}
(-1)^r  &  {\rm If} \; n = p_1 \dots p_r\\
0 & {\rm If} \; \exists  \, p, \; p^2  | n \\ 
\end{array}
\right. \ . 
\label{p19}
\eeq
In this expression $p_1, \dots, p_r$ are different prime numbers. Integer numbers  for which 
$\mu(n) \neq 0$ are called square free. The Moebius function has been used to construct
an ideal  gas of {\em primons}  with  fermionic statistics \cite{J90,S90}. In our model, $\mu(n)$ 
appears in order to  {\em amplify} the interference between  waves so that the Riemann zeros
are not  swept out in the semiclassical  limit and become  visible. 
%Perhaps these two uses of  $\mu(n)$ are related.  
In the  limit $k \rightarrow \infty$  one finds 
\beq
e^{ - i \Phi_\infty} \, R_\infty =  \varepsilon \sum_{n=1}^\infty   \frac{  \mu(n)}{n^{\sigma + i E}} =  \frac{\varepsilon }{\zeta(\sigma + i E) } 
, \quad {\rm if}  \left\{  \; 
\begin{array}{lc}
\sigma \geq 1   &  \\
\frac{1}{2} < \sigma < 1,   & {\rm RH} : {\rm True} \\
\end{array} \right.  . 
\label{p20}
\eeq
The case   $\sigma=1$ is equivalent to the Prime Number Theory, that was proved by Hadamard and de la Vall\'{e}e-Poussin
by showing that  $\zeta(1+ i t) \neq 0, \; \forall t \in \Rmath$ \cite{E74}.  Repeating the analysis performed
in the previous models, we find that for  $\sigma \geq 1$, the spectrum is given by  $\Rmath$. Indeed, 
for all values of $E$,  the norm of ${\bf A}_k$
approaches a constant value in the limit $k \rightarrow \infty$, and consequently the norm (\ref{p9}) 
diverges logarithmically. The same thing  occurs  for  $1/2 < \sigma < 1$, under the RH according to which 
this region will be  free of   {\em zeros}. 

We are then left with the model  with $\sigma = 1/2$ 
to provide a  spectral realization of the {\em zeros}. Let us first approach this case in  the 
 limit $\sigma \rightarrow 1/2$ when
$E = E_n$ is a {\em zero}, that is  $\zeta( 1/2 + i E_n) =0$. Expressing the zeta function on the critical line as 
$\zeta(1/2 + i E) = Z(E) e^{ - i \theta(E)}$, where $Z(E)$ and $\theta(E)$ are the  Riemann Siegel functions  \cite{E74}
one finds
\beq
\zeta( \sigma + i E_n) = - i (\sigma - 1/2) Z'(E_n) e^{ - i \theta(E_n)} + O((\sigma- 1/2)^2), \quad \sigma \rightarrow \frac{1}{2}^+ \ , 
\label{p21}
\eeq
where $Z'(E) = dZ(E)/dE$. For the sake of the argument we have assumed    that $E_n$ is a simple zero of $Z(E)$, a fact that is unknown to hold 
for all {\em zeros}.  If that is the case, the sign of the derivative of $Z(E)$ at the {\em zeros} satisfy the rule

\beq
{\rm sign} \, Z'(E_n) = (-1)^{ n +  \frac{1}{2}( 1 + {\rm sign} (n) )}
\label{p211}
\eeq
where the  positive {\em zeros} are labelled as $n =1,2, \dots,$ and the negative {\em zeros} as $n = -1, -2, \dots. $
Equation (\ref{p211}) can be derived from the continuity of $Z(E)$ and the fact that $Z(1/2) < 0$. 
Plugging (\ref{p21}) and (\ref{p211})  into (\ref{p20}) gives
\beq
e^{ - i \Phi_\infty} \, R_\infty 
  \simeq \frac{\varepsilon e^{ i (\theta(E_n) + \frac{\pi}{2})}  }{ ( \sigma - 1/2)  Z'(E_n)  } =  
  \frac{\varepsilon e^{ i (\theta(E_n) -  \pi ( n +  \frac{1}{2}{\rm sign} (n)))}  }{ ( \sigma - 1/2)  |Z'(E_n)|  }, 
\quad \sigma \rightarrow \frac{1}{2}^+ \ , 
\label{p22}
\eeq
that leads to  the choice 
\beq
R_\infty \sim   \frac{\varepsilon }{(\sigma - 1/2)  |Z'(E_n)|} >0, \qquad e^{ - i \Phi_\infty}  \sim  e^{ i (\theta(E_n) - \pi ( n + \frac{1}{2} {\rm sign} (n)))} . 
\label{p23}
\eeq
Hence  the limit  $\sigma \rightarrow 1/2$, implies   $R_\infty \rightarrow + \infty$, whereby   the amplitude
${\bf A}_\infty^T$ blows   up unless  the parameter $\vartheta$ satisfies 
\barray 
\cos( \Phi_\infty - \vartheta)  & = & 1 \rightarrow  \cos(   \theta(E_n) -  \pi ( n + \frac{1}{2} {\rm sign} (n))   + \vartheta) = 1, 
\label{p24}
\earray 
in which case ${\bf A}_\infty^T=0$, corresponding to   a normalizable  state.  This is an important result
that we shall derive more rigorously a bit later, but now let us discuss its implications. 

\begin{figure}[t]
\begin{center}
\includegraphics[height= 5.0 cm]{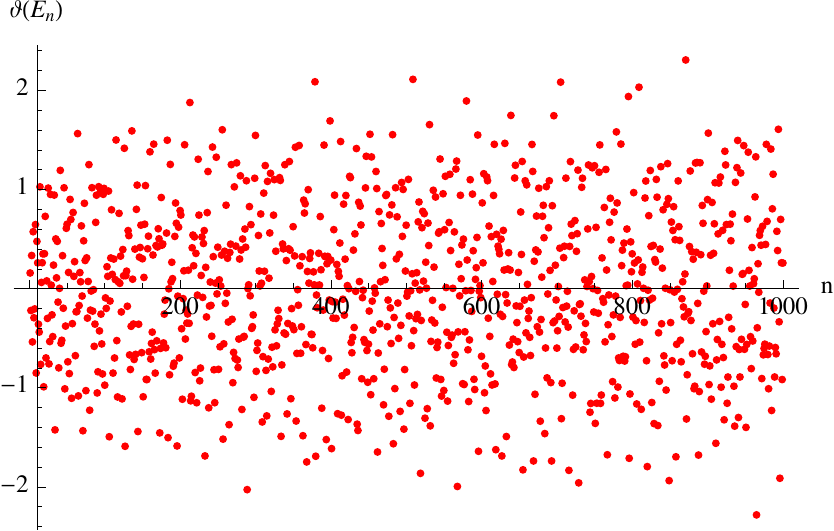} \hspace{1 cm}
\includegraphics[height= 5.0 cm]{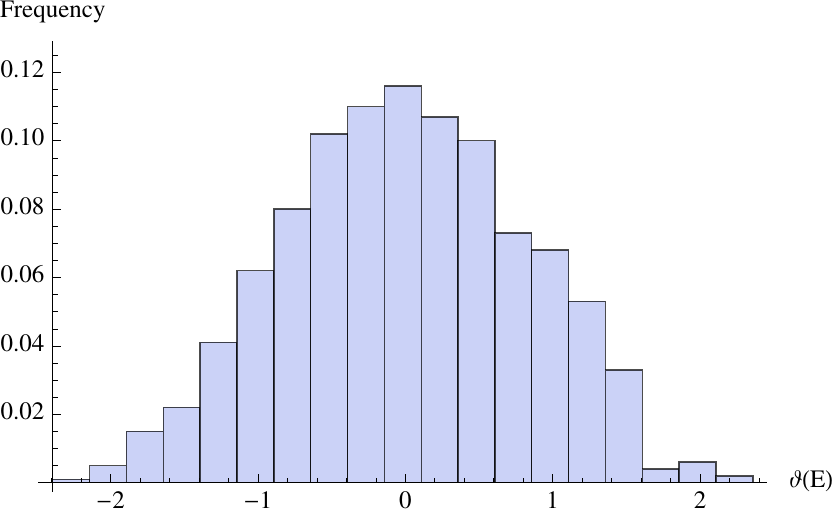}
\end{center}
\caption{Left: plot of $\vartheta(E_n) \in (- \pi, \pi]$ given in eq.(\ref{p241}) for the $10^3$ first positive Riemann  zeros. Right: histogram of   
$\vartheta(E_n)$. 
}
\label{thetan}
\end{figure}

\begin{itemize}

\item The value of $\vartheta(E_n)$ satisfying  eq.(\ref{p24})  can be written as  (with $E_n>0$)
\beq
\frac{\vartheta(E_n)}{\pi}  =     n + \frac{1}{2}  -  \frac{ \theta(E_n)}{\pi}   = n - \langle {\cal N}(E_n) \rangle + \frac{3}{2}, 
\label{p241}
\eeq
where $ \langle {\cal N}(E) \rangle$ is the average number of {\em zeros} in the range $[0, E]$  (see eq.(\ref{ze1})).  In the absence of fluctuations
the average would  be exact, that is  $\langle {\cal N}(E_n) \rangle = n$, whereby  $\vartheta(E_n) = 3 \pi/2, \; \forall n >0$. Hence 
a single value of $\vartheta$ would  work for  all the {\em zeros}. But the existence of fluctuations make things more interesting. To 
{\em hear} a given {\em zero} \cite{B12} one has to fine tune  $\vartheta$ 
according to eq.(\ref{p241}), which pass to  depend  on the  phase of the zeta function, $\theta(E_n)$.
One thus obtains that  the Riemann zeros and the phase of the zeta function  both  acquire 
a physical meaning in the model.

\item Berry \cite{B86},  and Badhuri et al.  \cite{BKL95} have argued that a better  approximation to the average {\em zeros} 
 is obtained if 
\beq
\cos \theta(E_n) \simeq  0 \rightarrow \frac{\theta(E_n)}{ \pi}  \simeq n + \frac{1}{2}, \quad n=1,2, \dots. 
\label{p242}
\eeq
This result can be visualized by plotting the real and imaginary parts of $\zeta(1/2 + i t)$ in the complex plane
for a large interval of $t$. One obtains a collection of loops that cut the real axis at the Gram points where
$\sin \theta(t) =0$, and that just before crossing the origin at $t = E_n$, the loop  cuts  the imaginary axis, where $\cos \theta(t)=0$.
Badhuri et al. also show that $\theta(t)$ gives roughly the scattering phase  shift of a non relativistic particle 
in an inverted harmonic potential ($V(x) \propto  - x^2)$  with a hard wall at the origin. 
Now, replacing (\ref{p242}) into (\ref{p241}) gives that in average $\vartheta(E_n) \simeq 0 \, {\rm mod} (2 \pi)$. 
This result is confirmed in Fig. \ref{thetan} which shows $\vartheta(E_n)$ for the first $10^3$ {\em zeros}
together with an histogram. 
 
 \end{itemize}

So far we have  considered the value of $e^{- i \Phi_\infty} R_\infty$, using eq.(\ref{p20}).
To show  the  existence of discrete states for $\sigma = 1/2$, we need 
the finite sum giving  $e^{- i \Phi_k} R_k$, that can be computed using the Perron's  formula
 \cite{Apostol,Ba11}
\beq
{\sum_{1 \leq n \leq x}}^*  \;  \frac{  \mu(n)}{n^z} = \lim_{T \rightarrow \infty} \int_{c- i T}^{c + i T} \frac{d s }{ 2 \pi i}  \frac{1}{\zeta(s+z)}  \frac{x^s}{s}, 
\qquad c > 0,  \quad c >     1- \sigma, \quad \sigma = {\rm Re} \, z  
\label{p27}
\eeq
where  ${\sum }^*$ means that the last term in the sum must be multiplied by 1/2 
when $x$ is an integer. The integral (\ref{p27}) can be done by residue calculus \cite{Ba11}
\beq
 \lim_{T \rightarrow \infty} \int_{c- i T}^{c + i T} \frac{d s }{ 2 \pi i} F(s) 
 = \sum_{ {\rm Re} \, s_j < c } {\rm Res}_{s_j}   \,   F(s), \quad F(s) =   \frac{1}{\zeta(s+z)}  \frac{x^s}{s}, 
\label{p28}
\eeq
where the sum runs  over the  poles $s_j$  of $F(s)$ located to the left of the line of integration, that is 
${\rm Re} \, s_j < c$. For $z=0$, the   sum (\ref{p27})  is basically the Mertens functions $M(x) = \sum_{n =1}^x \mu(n)$
that plays an important role in  Number Theory \cite{E74}-\cite{D}. 
Here we are interested  in  the values  $z = \frac{1}{2} + i E$, with $E$ a real number. Eq.(\ref{p27}) 
imposes the condition   $c > 1 - {\rm Re} \, z = 1/2$. Hence the poles of $F(s)$  that contribute to (\ref{p28}) have their real 
part   smaller than $c$, that is ${\rm Re} \, s_j \leq 1/2$.  The origin $s=0$ is a simple pole if $\zeta(z) \neq 0$
and a multiple pole  if $\zeta(z)=0$. In the latter case we shall assume that $z$ is a simple zero so that the pole is double.  
The corresponding residues are given by 
\beq
{\rm Res}_{s=0}   \,   F(s) = \left\{
\begin{array}{ll}
1/\zeta(z)   & {\rm if} \, \zeta(z) \neq 0,  \\ 
\log(x)/\zeta'(z) & {\rm if} \, \zeta(z) = 0, \zeta'(z) = 0. \\ 
\end{array}
\right. 
\label{p29}
\eeq
The remaining poles of $F(s)$ come  from the zeros of $\zeta(s+z)$, except  the case $s =0$,
that is  included in (\ref{p29}).  The trivial zeros of $\zeta$ contribute with  the poles  $s_n = - 2 n - z \;
(n=1, 2, \dots)$, with residue
\beq
{\rm Res}_{s=- 2 n - z}   \,   F(s) =  \frac{ x^{ - 2 n - z} } { - ( 2 n + z) \zeta'( - 2 n )}, \qquad n =1, 2, \dots, \infty .  
\label{p30}
\eeq
The non trivial zeros of $\zeta$, denoted as $\rho_m$, contribute with the poles $s_m = \rho_m - z$
(note that ${\rm Re} \, s_m < 1/2$), with residue
 \beq
{\rm Res}_{s= \rho_m- z}   \,   F(s) =  \frac{ x^{ \rho_m  - z} } { ( \rho_m - z) \zeta'( \rho_m )}, \qquad \rho_m \neq z 
\label{p31}
\eeq
Collecting results we find
\barray 
{\sum_{1 \leq n \leq x}}^*  \;  \frac{  \mu(n)}{n^z} &  = &  \frac{1}{ \zeta(z) } + 
\sum_{\rho_m}  \frac{ x^{ \rho_m  - z} } { ( \rho_m - z) \zeta'( \rho_m )} + \sum_{n=1}^{\infty}  \frac{ x^{ - 2 n - z} } { - ( 2 n + z) \zeta'( - 2 n )}, 
\quad   {\rm if} \,  \zeta(z) \neq 0,  
\label{p32} \\ 
{\sum_{1 \leq n \leq x}}^*  \;  \frac{  \mu(n)}{n^z}  &  = &  \frac{\log x}{ \zeta'(z) } + 
\sum_{\rho_m \neq z }  \frac{ x^{ \rho_m  - z} } { ( \rho_m - z) \zeta'( \rho_m )} + \sum_{n=1}^{\infty}  \frac{ x^{ - 2 n - z} } { - ( 2 n + z) \zeta'( - 2 n )}, 
\quad   {\rm if} \,  \zeta(z) = 0, \zeta'(z) \neq 0,  
\nonumber 
\earray 
These equations are formally exact,  so it  would be  interesting to prove them and  find their  range of validity. 
 In what follows, we  shall  derive their  consequences. 
Let us recall that the LHS of (\ref{p32}) gives essentially $e^{ - i \Phi_k} R_k$, where $k$ is to be identified with  $x$. 
Neglecting for a while the summands in these expressions one finds
\barray 
e^{ - i \Phi_k(E)} R_k(E)  & \sim &   \frac{ \varepsilon     \,  e^{  i  \theta(E)  }}{Z(E)},  \qquad Z(E) \neq   0,    \label{p33}  \\ 
e^{ - i \Phi_k(E_n)} R_k(E_n)  & \sim &   \frac{ \varepsilon   \log k  \,  e^{  i ( \theta(E_n) + \frac{\pi}{2})}  }{Z'(E_n)},  \qquad Z(E_n)=  0, \quad   Z'(E_n) \neq   0.
\nonumber
\earray  
In the first   case $R_k(E)$ remains bounded in the limit $k \rightarrow \infty$, that would yield a state in the continuum.
In the second  case  we choose 

\beq
R_k(E_n)  \sim \frac{ \varepsilon \log k}{ |Z'(E_n)|} >0, \qquad e^{ - i \Phi_k(E_n)}  \sim  e^{ i (\theta(E_n) - \pi ( n + \frac{1}{2} {\rm sign} (n)))} , 
\quad k \rightarrow \infty
\label{p34}
\eeq
that can be compared with (\ref{p23}). Then imposing eq.(\ref{p24}), yields the asymptotic behavior of the norm of  ${\bf A}_k^T$ (see eq.(\ref{t7}))  
\beq
\langle {\bf A}_{k}^T   | {\bf A}_{k}^T  \rangle  \rightarrow    2   e^{- 2 R_k}  = k^{- 2 \varepsilon/|Z'(E_n)|}, \quad k \rightarrow \infty, 
\label{p35}
\eeq
and a wave function $\chi^T$ whose norm  given by  (\ref{p9})  
\beq
\langle \chi^T  | \chi^T   \rangle  \sim \frac{1}{2} 
\sum_{n=1}^\infty  \frac{1}{n}   \langle {\bf A}_{n}^T   | {\bf A}_{n}^T  \rangle \sim 
\sum_{n=1}^\infty  \frac{1}{n^{ 1 + 2 \varepsilon/|Z'(E_n)|  }} = \zeta \left( 1 + \frac{ 2 \varepsilon}{ |Z'(E_n)|} \right),   
\label{p36}
\eeq
which is finite for any $\varepsilon >0$ corresponding to a discrete  eigenstate with energy $E_n$.

\begin{figure}[t]
\begin{center}
\includegraphics[height= 4.5 cm]{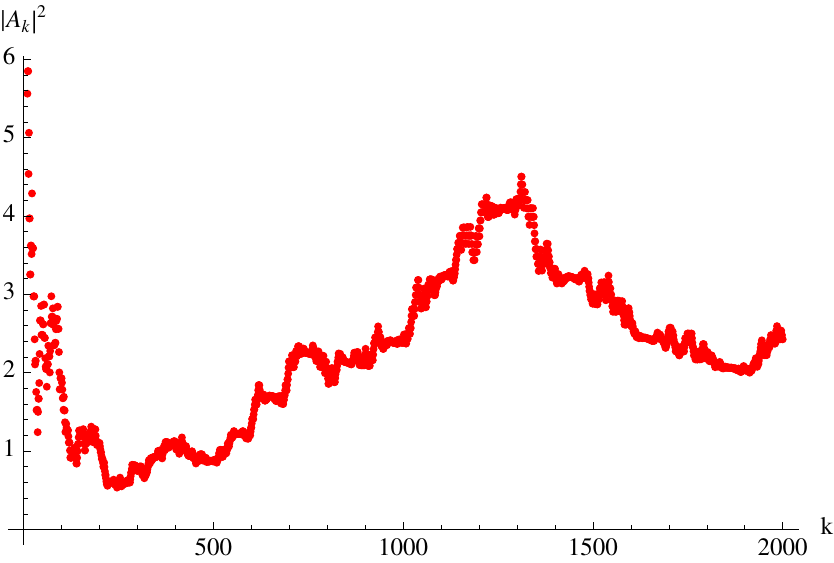} \hspace{1 cm}
\includegraphics[height= 4.5 cm]{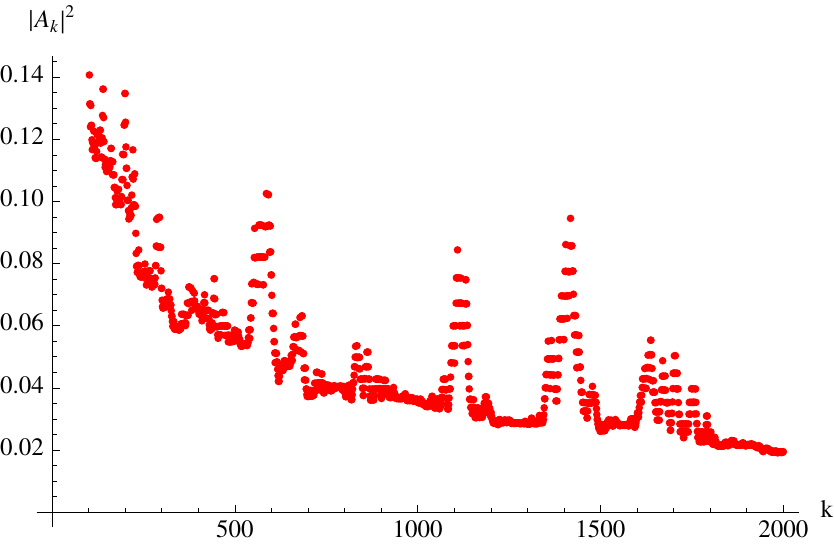} \hspace{1 cm}
\end{center}
\caption{Left: Plot of $|A_k|^2$ given in eq.(\ref{t7}) for $E=24$, with  $k=10, \dots, 2000$,
$\varepsilon =0.5, \vartheta = \pi$.   Right: $E_1=14.13..$, $k=10, \dots, 2000$, 
 $\varepsilon =0.25, \vartheta = \vartheta(E_n)$
}
\label{check3}
\end{figure}

Fig. \ref{check3}  shows $|{\bf A}_k|^2$ in the case where $E$ corresponds to a state in the continuum,
and for the first {\em zero} $E_1=14.13..$,  corresponding to a discrete state.  
In the latter case we took   $\vartheta = \vartheta(E_1)$, that guarantees that  the norm converges to zero except for some jumps. 
For other values like $\vartheta =\pi$ the norm increases  with $k$.  The same pattern
is observed for other {\em zeros}.  The values of $\vartheta(E_n)$ are mostly concentrated
around 0 (see Fig. \ref{thetan}). 
Finally,  we have computed the norm of the vector $\chi^T$, using
eq.(\ref{p9}) observing that,  for  the  {\em zeros},  it converges to a finite value with the choice
(\ref{p241}), and diverges in the remaining cases. These results are in agreement with eq.(\ref{p36}).

Let us now return to the summands in eq.(\ref{p32}) that we neglected in the previous computation.
The last term corresponding to the trivial {\em zeros} quickly converges to 0 as $x=k  \rightarrow \infty$.  The 
term associated to  non trivial {\em zeros} on the critical line, $\rho_m = 1/2 + i E_m, \; E_m \in \Rmath$,
oscillates as  $x^{ \rho_m - z}= e^{ i ( E_m - E_n) \log x}$, and we expect that it  gives  subleading contributions to the main term that goes with 
$\log x$. Finally,  a {\em zero}  off the critical line, say   $\rho_m = 1/2 +  E'_m + i E_m, \; E_m, E'_m >0 \in \Rmath$, 
would give  a contribution $x^{ \rho_m - z}= x^{E'_m}  e^{ i ( E_m - E) \log x}$, that dominates over the
remaining terms, for all  values  of $E$, leading to 
\beq
R_k(E_n)  \sim \varepsilon C(\rho_m, E)\,  k^{E'_m} > 0,  \qquad e^{ - i \Phi_k(E_n)}  \sim  e^{ i (E_m - E) \log k + \alpha(\rho_m, E))} , \quad k \rightarrow \infty
\label{p37}
\eeq
where $C(\rho_m,E)$ and $\alpha(\rho_m, E)$ do not depend on $k$.  The expression of $R_k(E)$ diverges when $k \rightarrow \infty$,
but  unlike the case of (\ref{p34}), the phase $\Phi_k(E)$ cannot be fixed to a value that cancels the divergent term $e^{2 R_k}$ 
in the norm (\ref{t7}). Hence $|{\bf A}_k^T|$, would  grow typically as ${\rm exp} ( \varepsilon  C k^{E'_m})$, so that  the wave function 
$\chi^T$ will not be normalizable even in  the continuum sense.  
This occurs for any value of $E$, so we arrive at the paradoxical conclusion   that 
a {\em zero} off the critical line implies that the Hamiltonian $H_R$ does not have eigenstates !!  That's not certainly the
case because $H_R$ is a well defined self-adjoint operator, so we must conclude that off critical {\em zeros} do not exist.
This result seems to provide a proof of the Riemann hypothesis,  but one must be very cautious since it relies on several unproven 
assumptions.   

\vspace{0.2cm}

{\bf Comments:}

\begin{itemize}

\item In the spectral realization proposed above there cannot exist  {\em zeros} outside the critical line 
in the form of resonances. As explained above,  their  presence leads to the non existence of eigenvectors 
 of the Hamiltonian $H_R$. 

\item von Neumann and Wigner showed that in ordinary Quantum Mechanics it is possible to have a bound state  immersed
in the continuum \cite{VW29,GP90}. They used a potential  that decays as $1/r$
with oscillations that trap the particle thanks to interference effects. There is a general class
of models with this property  \cite{S69}-\cite{CH02}, and they all require a fine tuning of couplings.
That this phenomena may happen for the Riemann zeros was suggested in \cite{S07a} using   a $xp$
model with non local interactions. 

\item The delta function  potential needed to reproduce  the {\em zeros}
 depends on the Moebius function  $\mu(n)$ that  exhibits an almost random behavior. 
This result is reminiscent of the fractal structure of the quantum mechanical potentials built to reproduce  the lowest values of  prime numbers 
and the Riemann zeros. The latter potentials were built from the smooth ones obtained by Mussardo, for the primes  \cite{M97}, 
and Wu and Sprung, for  the {\em zeros} \cite{WS},  and have a fractal dimension around 2 and 1.5 respectively  \cite{R95}-\cite{SBH08}.

\end{itemize}

\subsubsection{Dirichlet model}

We shall briefly describe how the previous results can be generalized to   Dirichlet $L_\chi-$functions. 
These are analytic functions constructed from a Dirichlet character $\chi$ as  \cite{D}
\beq 
L_\chi(s) = \sum_{n=1}^\infty \frac{ \chi(n)}{n^s} =  \prod_{p} \frac{ 1}{ 1 - \chi(p) p^{-s}} 
\qquad \Re \, s > 1, 
\label{d1}
\eeq
where the  product  runs  over the prime numbers (Euler formula). A character $\chi$ of modulus $q$, is 
an arithmetic function satisfying $\chi(m n ) = \chi(m) \chi(n), \chi(1) =1, \chi(m + q n) = \chi(m), \chi(n) =0$ if $q$ divides $n$ \cite{D}. 
One has that $|\chi(n)| =1$ or 0, and if  $\chi(n)= \pm 1, 0$,  the character is called real. 
The number of characters with modulus $q$ is given by the Euler totient function $\phi(q)$,
that counts  the number of coprime divisors of $q$.
For $q=1$, the character is $\chi_1(n) = 1, \; \forall  n $, that corresponds to the Riemann zeta function $\zeta(s)$. 
Primitive characters are those that cannot be written as products of characters of smaller modulus $q$, and 
can be  classified into even or  odd if $\chi(-1) = 1, -1$ respectively.

The $L$-functions associated to primitive characters satisfy the  functional relation 
\beq
\Lambda_\chi(s) \equiv \left( \frac{\pi}{q} \right)^{ - ( s + a_\chi)/2} \Gamma \left( \frac{ s + a_\chi}{2} \right)  L_\chi(s) 
=  e^{i \epsilon_\chi} \, \Lambda_{\bar{\chi}}(1-s), \quad e^{i \epsilon_\chi} = i^{ - a_\chi} \frac{ G_\chi}{\sqrt{q}}, 
\label{d5} 
\eeq 
where $a_\chi = 0  \, (1)$  for an even (odd) character,   $\bar{\chi}$ 
is the character conjugate to $\chi$, e.g.  $\bar{\chi}(n) = \chi^*(n)$,  and $G_\chi$ is the Gauss sum 
\beq
 G_\chi = \sum_{n=1}^q \chi(n) e^{ 2 \pi i n/q},    \qquad |G_\chi|   = \sqrt{q} . 
 \label{d3}
\eeq
On the critical line $s = 1/2 + i t$, eq.(\ref{d5}) reads 
\beq
 \left( \frac{\pi}{q} \right)^{ - i t/2} \Gamma \left( \frac{ 1 + 2 a_\chi}{4} + \frac{ i t}{2}  \right)  L_\chi\left( \frac{1}{2} +  i t \right) 
= e^{i \epsilon_\chi} \left( \frac{\pi}{q} \right)^{  i t/2} \Gamma \left( \frac{ 1 + 2 a_\chi}{4} -  \frac{ i t}{2}  \right)
  L_{\bar{\chi}}\left( \frac{1}{2} -  i t \right).
\label{d6} 
\eeq 
For $t$ real it is useful to  parameterize  the $L$ functions as
\beq
t \in \Rmath \rightarrow   L_{\bar{\chi}}^*\left( \frac{1}{2} -  i t \right) = L_\chi \left( \frac{1}{2} +  i t \right) = Z_\chi(t) \,  e^{ - i \theta_\chi(t)} , 
 \label{d7}
 \eeq
where $Z_\chi(t)$ is real and   $e^{ i \theta_\chi(t)}$ is a phase that can be found from eq.(\ref{d6})
\beq
e^{ 2 i \theta_\chi(t)}  = e^{- i \epsilon_\chi} \,  \left( \frac{\pi}{q} \right)^{ - i t}
\frac{ \Gamma \left( \frac{ 1 + 2 a_\chi}{4} +  \frac{ i t}{2}  \right) }{\Gamma \left( \frac{ 1 + 2 a_\chi}{4} -  \frac{ i t}{2}  \right) } , 
\quad e^{ 2 i ( \theta_\chi(t) +\theta_\chi(-t))  } =  e^{- i \epsilon_\chi} . 
\label{d8}
\eeq
$\theta_\chi(t)$ coincides with $\theta(t)$ when $\chi$ is the identity character. 
For real characters,  the zeros of $L_\chi$,  appear  symmetrically, $L_\chi(1/2 + i t_n) =  L_\chi(1/2 - i t_n)=0$, that is reflected
in the properties,  $Z_\chi(-t)  = Z(t)$ and $\theta_\chi(-t) = - \theta_\chi(t)$. However for complex characters this symmetry is broken.

The Dirac  model associated to  $L_\chi$ will be  defined by the parameters 
\beq
\ell_n = n^{1/2} ,   \quad \varrho_n = \varepsilon \,  \frac{  \mu(n) \chi(n)}{n^\sigma} \; \; (\sigma >0), \quad n =1, \dots, \infty .
\label{d9}
\eeq
Note that one can deal  with  complex characters thanks to the existence of two types of mass terms, 
$\bar{\psi} \psi$ and $\bar{\psi} \gamma^5 \psi$. The characters $\chi(n)$  acquire 
a physical meaning related to the reflection coefficient of the $n^{\rm th}$ mirror. This provides
a unified framework to deal with  the whole
family of Dirichlet $L$-functions.  

We can repeat the  analysis done for the zeta function to find the discrete eigenenergies  in the spectrum. 
For example,  the asymptotic limit of $e^{ - i \Phi_k} \, R_k$ is given by  
\beq
e^{ - i \Phi_\infty} \, R_\infty =  \varepsilon \sum_{n=1}^\infty   \frac{  \mu(n) \chi(n)}{n^{\sigma + i E}} = \frac{ \varepsilon }{L_\chi(\sigma + i E)},
\label{d10}
\eeq
which shows that  the zeros of $L_\chi(1/2 + i E)$ appear as poles of $R_\infty$, that lead eventually to discrete states, 
provided $\vartheta$ is fine tuned appropriately. Assuming that the zeros of $L_\chi(1/2 + i E)$ are
simple, the value of $\vartheta(E_n)$ for which $E_n$ is a discrete state is given by
\beq
\frac{\vartheta(E_n)}{\pi}  =     n + \frac{1}{2} \left( 1 + b_\chi + {\rm sign} (n) \right)   -  \frac{ \theta(E_n)}{\pi} , 
\qquad b_\chi = {\rm sign} \, Z_\chi(1/2).
\label{d11}
\eeq
For the zeta function, $b_\chi=-1$, eq.(\ref{p241}) is recovered from (\ref{d11}).  
We expect the {\em zeros} of the Dirichlet $L$ functions, associated
to primitive characters,  to  form the discrete spectrum of the corresponding Hamiltonians. 
That would amount to a proof of the Generalized Riemann hypothesis.

\section{CPT  symmetries and  AZ  classes}

The statistical properties of the Riemann zeros, conjectured by Montgomery and confirmed
numerically  by Odlyzko, have been one of the main motivations  to search for a spectral origin of these numbers.
The conjecture is that the {\em zeros} satisfy, locally, the GUE law, which implies 
that the  Riemann dynamics breaks the time reversal symmetry. Deviations from  the GUE law
were later on  identified by Berry and collaborators, as  a trace  of the semiclassical origin of the {\em zeros}
and a breakdown of universality. It is thus of great interest to study  the discrete  CPT 
symmetries  and RMT universality  classes of  the Hamiltonians  discussed in  previous sections. 

The action of the  time reversal symmetry (${\cal T}$),
charge conjugation or particle-hole symmetry (${\cal C}$),  and parity or chirality (${\cal P}$)
on a vector $\psi$ are defined as \cite{BL1,BL2}
\beq
\psi^{\cal T} = T \, \psi^*, \qquad \psi^{\cal C} = C \, \psi^*, \qquad \psi^{\cal P} = P \, \psi, 
\label{s1}
\eeq
where $T, C,P$ are unitary matrices, i.e. $T T^\dagger = C C^\dagger = P P^\dagger= {\bf 1}$ and 
$\psi^*$ is the complex conjugate of $\psi$. Here  $\psi$  is the  column vector
formed by the coefficients of a pure state of a Hilbert space in an orthonormal basis. 
${\cal T}$ and ${\cal C}$  are antiunitary transformations and  ${\cal P}$ is unitary.
A Hamiltonian $H$ has 
${\cal T, C}$ or ${\cal P}$ symmetry if it satisfies the conditions
\beq 
H^{\cal T}  =  T \, H^* \, T^\dagger   = H,  \qquad  H^{\cal C}  =  C \, H^t  \, C^\dagger = - H, \qquad 
H^{\cal P}  =  P \, H \, P^\dagger = - H. 
\label{s2}
\eeq
Since $H$ is hermitean, $H^* = H^t$, then  the ${\cal T}$ symmetry becomes 
$T \, H^t  \, T^\dagger   = H$. Notice  that ${\cal P}$ is the product of ${\cal T}$
and ${\cal C}$, and one can choose $P = C T^\dagger$. 
There is a basis where  $T$ and $C$ are real, and symmetric or antisymmetric matrices, i.e.
$T^t = \pm T$ and $C^t = \pm C$, in which case unitarity implies $T^2 = \pm {\bf 1}$
and $C^2 = \pm {\bf 1}$. If a symmetry  is broken, say ${\cal T}$, one writes  $T^2=0$. Counting all the possibilities one arrives to 
ten symmetry  classes,   as found by  Altland and Zirnbauer (AZ), 
that include  the classical Wigner-Dyson gaussian ensembles: GOE, GUE, and GSE 
as well as their chiral versions  chGOE, chGUE, chGSE  \cite{AZ}. 
Among the 10 AZ classes there are 4 that break time reversal symmetry, which 
are  the  candidates to describe  the {\em zeros} of $\zeta$ and other $L$-functions (see Table 1).
\begin{center}
\begin{tabular}{|c|c|c|c|c|}
\hline 
AZ   & $T^2$  & $C^2$  & $P^2$  & Top  \\
 \hline 
A  & 0 & 0 & 0  & 0  \\
AIII   & 0 & 0 & 1  & $\Zmath$  \\
D  & 0 & 1 & 0   & $\Zmath_2$  \\
C   & 0 & -1 & 0   & 0 \\
 \hline
 \end{tabular}
 
 \vspace{0.2cm} 
 Table 1.- The AZ classes where  the time reversal symmetry is broken. Column "Top"
 denotes the topological invariants of the class in one spatial dimension.  
 \end{center} 

Let us review briefly the main properties of the classes of Table 1 and their relations with the problem at hand. 
Class $A$ characterizes Hamiltonians of the form $H= A$, where $A$ is an hermitean matrix, with
no further conditions placed on it. The statistical properties of random matrices of this form are  described by GUE. As shown
in Table 1, all the CPT symmetries are broken and there is no topological invariant in 1D. 

Class D characterizes Hamiltonians of the form $H= A$, 
where $A$ is imaginary  and antisymmetric.  The  eigenvalues of  $A$ appear in pairs $\{ E_n, -E_n \}$,
and  if the dimension of the matrix $A$ is odd there is a zero eigenvalue. 
 The Berry-Keating Hamiltonian $H = (x \hat{p} + \hat{p} x)/2$
belongs to this class, since here $T=C=1$ and $H^t = - H$ \cite{S11b}.  The
Hamiltonian $H = \sqrt{x} ( \hat{p} + \ell_p^2 \hat{p}^{-1} ) \sqrt{x}$ also belongs to class $D$  provided 
the parameter $\vartheta$,  that characterizes its self-adjoint extensions,  is $0$ or $\pi$.  The latter choices 
ensure that  the eigenvalues of $H$  come  in pairs $\{ E_n, -E_n \}$,
and that if $\vartheta =0$, then $E_0=0$  is eigenvalue \cite{SL11,S12}. We can thus identify $\vartheta/\pi=0, 1$
as the  $\Zmath_2 =\{ 0, 1\}$ topological invariant of class $D$ in 1D. 

Class C characterizes  Hamiltonians of the form $H = A  + \vec{\sigma} \cdot \vec{S}$, where  $A^t = A^* = -A$,  $ \vec{S}^t= \vec{S}^* = \vec{S}$
and  $\vec{\sigma}$ are the Pauli matrices that act in an additional 2 dimensional Hilbert space,  that can be seen as
a spin 1/2  \cite{S11b}.  Here $C = i \sigma^y$, so $T^2=-1$. The eigenvalues of $H$  come in pairs $\{ E_n, -E_n \}$, as in class $D$. 
Srednicki proposed recently that class $C$ is associated to  the {\em zeros} of the Dirichlet
$L$-functions whose characters   
are real and even, that includes  the zeta function \cite{S11b}. He was led to this proposal by a conjecture due to  Katz and Sarnak 
\cite{KS99}  according to which these  $L$-functions form a {\em family} related by a sort of  symplectic symmetry,  and by the fact that 
 the spacings of their {\em zeros} agree asymptotically with the GUE distribution. 
 
Class AIII, is  a  chiral  version of GUE (chGUE) and characterizes Hamiltonians with  the block structure 
\beq
H = \left( 
\begin{array}{cc}
0 & A \\
A^\dagger  & 0 \\
\end{array}
\right), \qquad P =  \left( 
\begin{array}{cc}
1 & 0 \\
0  & -1 \\
\end{array}
\right), \qquad P H P^\dagger = - H, 
\label{s7}
\eeq
where $A$ is a complex matrix and $P$ the chiral operator. 
The eigenvalues of  $H$ come in pairs $\{ E_n, -E_n \}$. 
If $A$ is a matrix of dimension $N_+ \times N_-$, then the number of zeros eigenvalues is $|N_+ - N_-|$
that explains the  $\Zmath$ topological invariant of this class. Class chGUE, together with 
its  relatives chGOE and chGSE,  describe massless Dirac fermions and have been 
applied to study the QCD  Dirac operator,  partition functions, etc  \cite{VW}-\cite{FS02}. 

Let us next  study the symmetry classes of the Rindler Hamiltonians constructed  in previous sections.
First  of all, the massless Hamiltonian 
\beq 
H_R  = \sqrt{\rho} \,  \hat{p}_\rho \,  \sqrt{\rho} \,   \sigma^z, 
\label{s8}
\eeq
admits different representations  of the  $CPT$ symmetries \cite{BL2}. Indeed, $T$  and $P$ can be realized as $\sigma^x$ or $i \sigma^y$,
and  $C$ as $\sigma^z$ or ${\bf 1}$. Adding a mass term to (\ref{s8}) selects a particular realization, e.g.
\beq
H_R  = \sqrt{\rho} \,  \hat{p}_\rho \,  \sqrt{\rho}  \sigma^z + m \rho \sigma^x \rightarrow T = \sigma^x, \quad C = \sigma^z, \quad P = i \sigma^y. 
\label{s9}
\eeq
This Hamiltonian acts on the wave functions satisfying the BC (\ref{R50}). One can verify that this BC preserves 
${\cal T}$, for all values of $\vartheta$,  however the symmetry ${\cal C}$ is preserved only if $\vartheta =0, \pi$.  
Hence in these two cases, the Hamiltonian (\ref{s9}) belongs to class BDI (chGOE)  since $T^2= C^2 =1$. 
We mentioned in section II.E that the Hamiltonian $H = \sqrt{x} ( \hat{p} + \ell_p^2 \hat{p}^{-1} ) \sqrt{x}$( see eq. (\ref{R522}))
 has the same spectrum as (\ref{s9})
with the identifications (\ref{R523}).  We show above that the former Hamiltonian belongs to class D, while
we have found now that (\ref{s9}) belongs to  class chGOE.   There is no contradiction between these results.
In both cases the ${\cal C}$ symmetry explains  the pairing of energies $\{E_n, - E_n\}$, while the ${\cal T}$
symmetry appears from the doubling of degrees of freedom in the Dirac model. 

Finally, let us show that the Dirac model associated to the Riemann zeros
belongs to class AIII, with chiral operator  $P=  i \sigma^y$. First of all notice
that the matching conditions (\ref{2021}) are preserved by the action of $P$, because in this model $r'_n = r''_n=0$, and then
\beq
\chi(\ell_n^-) = L_n \, \chi(\ell_n^+) \rightarrow \chi^P(\ell_n^-) =  L_n  \,  \chi^P(\ell_n^+), \quad  L_n = P L_n P^\dagger . 
\label{s10}
\eeq
On the other hand,  if $\chi$ is the eigenfunction with energy $E_n$ it satisfies the BC (\ref{dm22}) 
\beq
- i e^{ i \vartheta(E_n)} \chi_-(\ell_1) = \chi_+(\ell_1) \rightarrow - i e^{-  i \vartheta(E_n)} \chi_-^P(\ell_1) = \chi_+^P(\ell_1) , 
\label{s11}
\eeq
where $\chi^P_\mp = \pm  \chi_\pm$. The function $\vartheta(E_n)$,  given in eq.(\ref{p241}) for $n>0$, 
satisfies that $- \vartheta(E_n) = \vartheta(-E_n)$  which,  together with the equation  $P H_R  P^\dagger = - H_R$, 
 implies that  if $\chi$ is an eigenstate with energy $E_n$, then 
 $\chi^P$ is an eigenstate with energy 
$- E_n$. Hence the pairing of energies is explained by the chiral symmetry and not by  the charge 
conjugation symmetry as the Hamiltonian (\ref{s9}). One can perform a change of basis that brings
the chiral operator $P$  and the Hamiltonian  (\ref{s8}) into the form (\ref{s7}), with $A= i \sqrt{\rho} \,  \hat{p}_\rho \,  \sqrt{\rho}$
that is a real and antisymmetric matrix in some orthonormal basis. 
The Hamiltonian we are dealing with is therefore a subclass within the 
class chGUE.

\section{Conclusions} 

We have proposed in this paper  that a combination of Quantum Mechanics and Relativity Theory 
is the key to the  spectral realization of the Riemann zeros. The old message of P\'olya and Hilbert
was that Quantum Mechanics should play a central role in this realization, an idea that is behind 
the successful  applications of  RMT,  and  Quantum Chaos,  to Number Theory.  What is new is that Einstein's theory of 
Relativity must  also be present.  The reason is that the  properties of accelerated objects, 
in Minkowski space-time,  can be used to encode and process arithmetic information. 
In some  sense, spacetime becomes an analogue computer, or simulator,  that allows  an accelerated  observer
to multiply numbers, and distinguish primes from composite,   by measuring the proper times 
of events involving massless particles.  In this way  the prime numbers acquire a classical  
relativistic meaning  associated to  {\em primitive}   trajectories of particles, along the lines
suggested  by the periodic orbit theory in Quantum Chaos.  

The combination of Quantum Mechanics and Relativity is of course  a Relativistic
Quantum Field Theory, that in our case   is the Dirac theory of fermions 
 in Rindler space-time, the latter being  the geometry associated  to moving 
observers.  The time evolution in this space is generated by the Rindler Hamiltonian
that coincides with the Berry-Keating Hamiltonian $\pm (x  \hat{p} +  \hat{p} x)/2$, each sign
corresponding to  the chirality of the fermion. This operator is also the generator
of dilations of the Rindler radial coordinate $x$, but in the presence of delta function potentials, 
associated to the prime numbers,  scale invariance is broken. This has the  dramatic effect
that the spectrum is a continuum, where the Riemann zeros are missing, in analogy
with the result found by  Connes in the adelic approach to the RH.
However, the moving observer can  fine tune  the phase factor of   the reflection 
of the fermion at the boundary, in such a way  that  a bound state appears with an  
energy given by  a Riemann zero.   The phase factor
at the boundary is given essentially by the phase of the zeta function at the corresponding 
{\em zero}.  This result is obtained in a limit where
the perturbation of the massless Dirac action by the delta potential is infinitesimally small, 
so that the bound state is  immersed in a  continuum of states.  This  weak coupling limit is  reminiscent
to the semiclassical limit that leads to the Gutzwiller formula for the fluctuations of the energy levels
in chaotic systems. The previous scenario leads  to a proof of the Riemann hypothesis
but more work is required to fully support this  claim. 

The present  work suggests the  possibility of an experimental observation of the Riemann zeros as energy levels. 
The system proposed here does not look realistic at the moment, but there are   theoretical  proposals
to realize the Rindler geometry using  cold atoms and optical lattices \cite{Latorre1,Javi}.  Another possible
route is to use  the Quantum Hall effect where $xp$ arises as an effective low-energy Hamiltonian \cite{ST08,St13}. 
On a more theoretical perspective, this  work points towards  a relation between 
Riemann zeros and  black holes, whose near horizon geometry
is in fact  Rindler \cite{G}.
The role played by the privileged observer in our construction reminds the brick 
wall model introduced by  't Hooft to regularize the horizon of a black hole \cite{H85}. 
Throughout this paper the prime numbers have been treated as classical objects. 
However, they turn  into quantum objects in the  Prime state 
formed by the superposition of primes in the  computational basis of a quantum computer \cite{LS1,LS2}.  
This poses the question  whether the Prime state could be created with  the same tools as the Riemann zeros become energy levels.

\vspace{1 cm} 

{\bf Acknowledgements.- }  I am very   grateful   for suggestions, and conversations to Jos\'{e} Ignacio
Latorre, Bel\'en Paredes, Paul Townsend,  Michael  Berry, Jon Keating,  Javier Rodr\'{i}guez-Laguna, 
Manuel Asorey,    Jose Luis Fern\'andez  Barb\'on,   Giuseppe Mussardo,  Andr\'e  LeClair,    Mark  Srednicki, 
Javier Molina, Luis Joaqu\'{\i}n Boya and  Miguel Angel Mart\'{\i}n-Delgado. 
This work has  been financed  by the Ministerio de  Ciencia e Innovaci\'on, 
Spain (grant FIS2012- 33642), Comunidad de Madrid (grant QUITEMAD) and the Severo Ochoa Program.

\appendix

\section{The harmonic model}

We first discuss some general   properties of the transfer matrix $T_k$ that  will be  used later on
 to derive the exact spectrum of the {\em harmonic}  model.
Let us consider the  transfer matrix (\ref{dm34}), with $r'_k =0$,  and the condition $|r_k| <1$.
It is easy to see that $T_k$ can be written as 
\beq
T_k =   \frac{1}{1 - r_k^2} 
\left( \begin{array}{cc}
r_k^2 + 1 &  2 r_k  \, \ell_k^{ - 2 i E } \\
 2 r_k  \ell_k^{2  i E} & r_k^2 + 1   \\
\end{array}
\right) =  e^{ - i  \varphi_k \sigma^z} \, e^{ g_k \sigma^x} \, e^{  i  \varphi_k  \sigma^z}, 
\label{ap1}
\eeq
where $e^{ \pm g_k}$ are the eigenvalues of $T_k$ (positive from the condition $|r_k| <1$)
\barray
e^{g_k} &  = &  \frac{1 + r_k}{ 1 - r_k}, \quad e^{-g_k} = \frac{1 - r_k}{ 1 + r_k}, 
\label{ap2} \\
{\rm sign}  \, g_k  & =  &  {\rm sign} \,  r_k, \quad  \quad |r_k| < 1, 
\nonumber 
\earray
and 
\beq
\varphi_k = E   \, \log \ell_k , \qquad k=1, \dots, \infty . 
\label{ap3}
\eeq
Replacing  (\ref{ap1}) into  (\ref{dm31}) gives the recursion relation  
\beq
| {\bf A}_{k-1}  \rangle   =  e^{ - i  \varphi_k \sigma^z} \, e^{ g_k \sigma^x} \, e^{  i  \varphi_k  \sigma^z} \,  | {\bf A}_{k} \rangle,    \qquad  k \geq 2.
\label{ap4}
\eeq
We now define the vector
\beq
| {\bf \hat{A}}_{k}  \rangle   = e^{  i  \varphi_k  \sigma^z} \,  | {\bf A}_{k} \rangle, \qquad  k \geq 1, 
\label{ap5}
\eeq
that satisfies
\beq
| {\bf \hat{A}}_{k-1}  \rangle   =  e^{ - i  \Delta _k \sigma^z} \, e^{ g_k \sigma^x} \, | {\bf  \hat{A}}_{k} \rangle,    \qquad  k \geq 2, 
\label{ap6}
\eeq
where
\beq
\Delta_k = \varphi_{k}- \varphi_{k-1} = E  \,  \log \frac{ \ell_k}{ \ell_{k-1}}, \qquad k \geq  2. 
\label{ap7}
\eeq
Notice that
\beq
| {\bf \hat{A}}_{1}  \rangle   =   | {\bf A}_{1} (\vartheta) \rangle.
\label{ap8}
\eeq
Eq.(\ref{ap6}),  gives an alternative way to solve the eigenvalue problem that admits an interesting
physical interpretation as the evolution of a kicked rotator with spin 1/2 \cite{St99}. In this interpretation
the term $e^{ - i \Delta_k \sigma^z}$ represents the rotation  of the spin around the $z$-axis,  
with an  energy $E$ during a time elapse $\log (\ell_k / \ell_{k-1})$, after which the spin is kicked with an imaginary magnetic field with strength $g_k$
along the $x$-direction.   Kicked rotators of this sort are currently employed  to  analyze  quantum chaos in simple situations \cite{St99}.

\begin{figure}[t]
\begin{center}
\includegraphics[height= 6.0 cm]{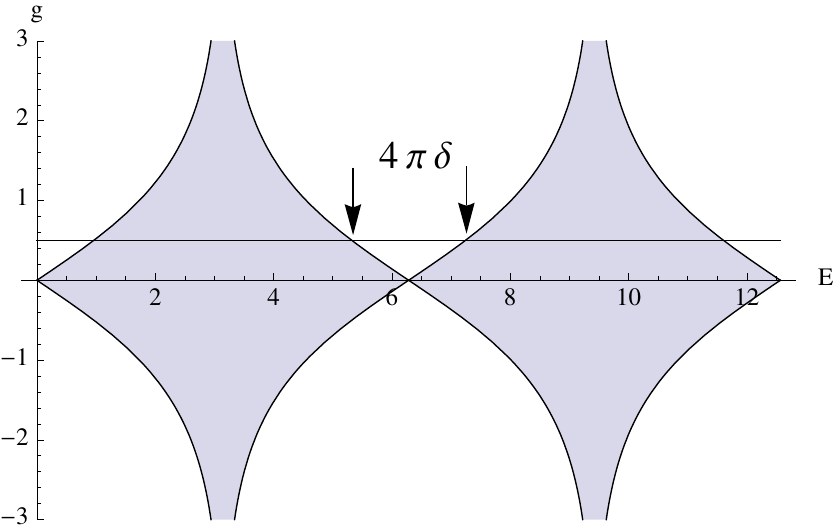} \hspace{1 cm}
\end{center}
\caption{ The region in shadow represents the values of $E$ and $g$ where the matrix $S$ is elliptic,
that corresponds to states in the continuum
(see eq.(\ref{ap15})).  In  the white region the matrix $S$ is hyperbolic, and contains  the discrete spectrum whose location 
depend on  $g$ and $\vartheta$  by eq.(\ref{ap14}). For $\vartheta=0, \pi$, the eigenstates are at $E_n = 2 \pi n$ for all values of $g$
(see eq.(\ref{ap13})). The  horizontal line corresponds to a value of the coupling constant $g$. Its intersection with the shadow regions
give the continuum spectrum that form the bands  (\ref{ap16}) separated by a gap $4 \pi \delta$. }
\label{gdelta}
\end{figure}

Let us next  consider the {\em harmonic}  model defined by eq.(\ref{p1}), that according to eqs.(\ref{ap2}, \ref{ap7}) 
correspond to constant values of $g_k = g$ and  $\Delta_k = \Delta$ 
\beq
\ell_n = e^{n/2}, \quad \varrho_n = r_n = \varepsilon \longrightarrow e^g =  \frac{1 + \varepsilon}{ 1 - \varepsilon}, \quad
\Delta = \frac{E}{2} . 
\label{ap9}
\eeq
In this case  the recursion relation (\ref{ap6}) can be  iterated yielding 
\beq
| {\bf \hat{A}}_{k+1}  \rangle   = \, S^k   \, | {\bf  {A}}_{1}(\vartheta) \rangle,   \qquad  k \geq 0, 
\label{ap10}
\eeq
where the matrix $S$ is given by 
\beq
 S  = e^{ - g \sigma^x} e^{ i E \sigma^z/2} = \left( 
 \begin{array}{cc}
 \cosh g \;  e^{ i E/2} & -  \sinh g  \; e^{ -i E/2}  \\
 -  \sinh g \;  e^{ i E/2} &  \cosh g  \; e^{- i E/2} \\
 \end{array}
 \right) . 
\label{ap11}
\eeq
Making the replacement $E \rightarrow E  + 2 \pi$, the matrix $S$ changes by an overall sign, which in turn
implies that the  spectrum has a $2 \pi$  periodicity.  Let us consider the case where  $E = 2 \pi n$, that yields
\beq 
S = (-1)^n \, e^{ - g \sigma^x} \longrightarrow | {\bf \hat{A}}_{k+1}  \rangle   = \, (-1)^{n k} \, e^{ - g k  \sigma^x} \,
\left( \begin{array}{c}
1 \\
e^{ i \vartheta} \\
\end{array}
\right), 
\label{ap12}
\eeq
which is convergent only in the following  two cases
\barray 
\vartheta = 0,  & g >0 &   \longrightarrow | {\bf \hat{A}}_{k+1}  \rangle   = \, (-1)^{n k} \, e^{ - g k } \,
\left( \begin{array}{c}
1 \\
1\\
\end{array}
\right) , 
\label{ap13}  \\
\vartheta = \pi,  & g  < 0 &   \longrightarrow | {\bf \hat{A}}_{k+1}  \rangle   = \, (-1)^{n k} \, e^{ - |g|  k } \,
\left( \begin{array}{c}
1 \\
-1\\
\end{array}
\right) . \nonumber 
\earray 
These are normalizable eigenstates for all values of the coupling constant $g$, even in  the limit $g \simeq 2 \varepsilon \rightarrow 0$. 
Normalizable states for  $\vartheta \neq 0, \pi$ are  also possible. The relation between the energy, $E$,  $g$ and $\vartheta$ is given by 
\beq
\tan (E/2) =  \frac{ \sin \vartheta}{Ê\cos{\vartheta} + \coth g},  \qquad  1 + \cos \vartheta \, \coth g  >0, 
\label{ap14}
\eeq
that in the limit $g \rightarrow 0$ gives the discrete eigenvalues $E \rightarrow  2 \pi n$. Plugging $\vartheta =0, \pi$ in (\ref{ap14}) we recover the 
cases (\ref{ap13}). 
 The continuum spectrum of the model 
can be derived from the condition that $S$   is an elliptic matrix (note that $\det \, S=1$)
\beq
|\Tr \, S = 2 \cosh g \, \cos E/2 | < 2  \longleftrightarrow |\sin E/2 | > \tanh g , 
\label{ap15}
\eeq
and consists of energy bands separated by a gap $4 \pi \delta$ that are shown in Fig. \ref{gdelta} 
\beq
{\rm Spec}_c = \cup_{n=- \infty}^\infty  2 \pi
[ n + \delta, n + 1 - \delta], \qquad \sin ( \pi \delta ) = \tanh g = \frac{ 2 \varepsilon}{ 1 + \varepsilon^2}.
\label{ap16}
\eeq
In the limit $g\rightarrow 0$, the gap closes, and  the discrete energy states $E_n = 2\pi n$ 
become immersed in the continuum $\Rmath$.

\end{document}